\documentstyle[psfig]{mn}
\def\mag{\hbox{$^{\rm m}$}}

\def\degr{\hbox{$^\circ$}}

\def\arcsec{\hbox{$^{\prime\prime}$}}

\title[The Stellar Composition of the Star Formation Region CMa R1]
  {The Stellar Composition of the Star Formation Region CMa R1 -- 
   III. A new outburst of the Be star component in Z~CMa
        \thanks{Based on observations made with the European Southern Observatory
        telescopes obtained from the ESO/ST-ECF Science Archive Facility,  
        with the William Herschel Telescope operated on 
        the island of La Palma by the Isaac Newton Group in the Spanish 
        Observatorio del Roque de los Muchachos of the Instituto de 
        Astrof\'{\i}sica de Canarias, with the BTA telescope of the Special 
        Astrophysical Observatory, Nizhnii-Arkhyz, Russia, and with 
        the Alfred Jensch telescope of the Th\"uringer Landessternwarte, 
        Germany.}}
\author[M.E. van den Ancker et al.]
  {M.E. van den Ancker,$^1$\thanks{Visiting Astronomer at the Infrared Telescope Facility, 
       which is operated by the University of Hawaii under Cooperative Agreement 
       no. NCC 5-538 with the National Aeronautics and Space Administration, Office of 
       Space Science, Planetary Astronomy Program.} P.F.C. Blondel,$^2$
   H.R.E. Tjin A Djie,$^2$ K.N. Grankin,$^3$\newauthor
   O.V. Ezhkova,$^{3,4}$ V.S. Shevchenko,$^{3,5}$ E. Guenther$^6$ 
   and B. Acke$^7$\thanks{E-mail contact: Herman Tjin A Djie (herman@science.uva.nl)}\\ \\
   $^1$European Southern Observatory, Karl-Schwarzschild-Strasse 2, 
       D-85748 Garching bei M\"unchen, Germany\\
   $^2$Astronomical Institute ``Anton Pannekoek'', University of Amsterdam,
       Kruislaan 403, 1098 SJ Amsterdam, The Netherlands\\
   $^3$Astronomical Institue of the Academy of Sciences of Uzbekistan,
       Astromicheskaya 33, Tashkent 700052 Uzbekistan\\
   $^4$Sternberg Astronomical Institute, Moscow State University,
       Universitetskii Prospect 13, RU - 119899 Moscow, Russia\\
   $^5$Deceased March 2000\\
   $^6$Th\"uringer Landessternwarte Tautenburg, Sternwarte 5, 
       D-07778 Tautenburg, Germany\\
   $^7$Astronomical Institute, Katholieke Universiteit Leuven,
       Celestijnenlaan 200B, B-3001 Heverlee, Belgium}
\date{Accepted $<$date$>$.
      Received $<$date$>$}
 
\pagerange{\pageref{firstpage}--\pageref{lastpage}}
\pubyear{2003}
 
\begin{document}
\maketitle
\label{firstpage}

\begin{abstract}
We report on a recent event in which, after more than a decade of 
slowly fading, the visual brightness of the massive young binary 
Z~CMa suddenly started to rise by about 1 magnitude in December 1999, 
followed by a rapid decline to its previous brightness over the 
next six months.  This behaviour is similar to that exhibited by 
this system around its eruption in February 1987.  
A comparison of the intrinsic luminosities of the 
system with recent evolutionary calculations shows that Z~CMa may consist 
of a 16 M$_\odot$ B0 IIIe primary star and a $\sim$3 M$_\odot$ FUor 
secondary with a common age of $\sim$ 3 $\times$ 10$^{5}$ yr.  We also 
compare new high-resolution spectra obtained in Jan. and Feb. 2000, 
during the recent rise in brightness, with archive data from 1991 and 1996.  
The spectra are rich in emission 
lines, which originate from the envelope of the early B-type primary star.  
The strength of these emission lines increased strongly with the brightness 
of Z~CMa.  We interpret the collected spectral data in terms of an accretion 
disc with atmosphere around the Herbig B0e component of Z~CMa, which has 
expanded during the outbursts of 1987 and 2000.  A high resolution
profile of the 6300\,\AA\ [O\,{\sc i}] emission line, obtained by us in March
2002 shows an increase in flux and a prominent blue shoulder to the feature 
extending to $\sim$ $-$700~km~s$^{-1}$, which was much fainter in the 
pre-outburst spectra.  We propose that this change in profile is a result of 
a strong change in the collimation of a jet, as a result of the outburst 
at the start of this century.
\end{abstract}


\section{Introduction}
   Ever since the discovery of bright emission lines in its 
spectrum by Merrill (1927), Z~CMa (HD~53179, IRAS 07013$-$1128) 
has been one of the most frequently studied massive young stars. 
The brightness of the system is highly variable, varying irregularly 
between its quiescent state at $V$ $\approx$ 11 to its active state 
at $V$ $\approx$ 9 (Covino et al. 1984; Hessman et al. 1991).  
Although the photometric and spectroscopic properties of Z~CMa show 
some similarity with those of the FUors (e.g. FU Ori and V1057 Cyg), 
its spectral differences with these objects, especially in the 
presence of many narrow emission lines, are remarkable. Nevertheless, 
in its normal brightness phases ($V$ $>$ 9.6\mag) the absorption 
character of the spectrum is dominant and, similar to the FUors, 
the spectrum can be interpreted as that of an optically thick
accretion disc (Hartmann et al. 1989; Welty et al. 1992).

  In Feb. 1987 a large rise in brightness of Z~CMa was accompanied
by a rise in emission flux in the Balmer lines and Ca\,{\sc ii}\,K 
and with the transition of many metal lines of Fe\,{\sc ii}, Cr\,{\sc ii}
and Ti\,{\sc ii} from absorption into emission (Hessman et al. 1991). 
These authors tried to
explain the absorption and emission spectral components in terms of a
common source, a variable accretion disc. However, it proved 
difficult to find a simple model for such a source that would be able 
to reproduce the observations.  A key ingredient in solving this mystery 
was discovered in 1989, when near-infrared speckle interferometry revealed 
that Z~CMa is a double star with a separation of 0.1\arcsec 
(Koresko et al. 1991;
Haas et al. 1993). These authors showed that the South-Eastern component 
of Z~CMa dominates the visual and UV parts of the SED and is associated 
with a FUor-like system, whereas the North-Western component is a 
powerful infrared source, which is the primary object of this double star 
system.  The separation of Z~CMa into two components 
has since been confirmed by speckle masking observations in the 
visual ($R$-band) by Barth et al. (1994) and by Thi\'ebaut et al. (1995).
In addition it was discovered by Whitney et al. (1993) that the
emission lines in the visual part of the spectrum of Nov. 1991 were 
polarized, in contrast to the continuum of
the spectrum. This supports the idea that these emission lines are 
contributed by scattered light from the primary source of the system.

The analogy of this spectrum with that of the early type Herbig Be 
star MWC 1080 suggests that the source of the emission lines is an 
early type Be star inside a large cocoon. The radiation of this Be star
escapes by scattering from the dust walls of a cavity in the thick cocoon 
envelope. 
Millan-Gabet \& Monnier (2002) succesfully imaged the direct 
circumstellar environment of Z~CMa at 1.25~$\mu$m, and found a small 
($\sim$ 1\arcsec) linear feature.  They interpreted this feature as 
light scattered off the walls of a jet-blown cavity, confirming the 
above picture.  The cavity could be formed by the extended (over 3.6 pc) 
optical jet discovered by Poetzel et al. (1989).  This jet is also 
seen as a bipolar outflow in CO (Evans et al. 1994).  More
recent spatially resolved optical spectroscopy by Bailey (1998) and 
Garcia et al. (1999) has confirmed that the extended outflow originates
from the infrared primary component of Z~CMa and that the optical emission
line spectrum is associated with the same component.  

During Nov./Dec. 1999, observers of the ROTOR photometric monitoring 
program at Mt. Maidanak Observatory (Uzbekistan) found that after seven 
years of low brightness ($V$ $\sim$ 10.3\mag) the brightness of Z~CMa 
rose sharply (within 40 days) by more than 1 magnitude.  We obtained 
new high-resolution spectra of Z~CMa shortly after this new outburst.  
These spectra show intrinsic changes in line profiles and 
equivalent widths with respect to spectra obtained 
in 1991 and 1996.  In this paper we discuss the spectral changes of 
Z~CMa over the last ten years and interpret these changes as due to 
changes in the size of an accretion disk, associated with the 
Herbig B0e primary.

From $UBVRI$ photometry of Z~CMa obtained in Nov. 1991 and Feb. 1987 
(Sect. 3.2) we confirm the spectral classification of the star in the 
primary source and derive its radius and luminosity for spectral type 
B0III. Although the photometric data of Z~CMa suggest that
the variability of the contributions from both binary components is
for a large part due to irregular variations in their circumstellar 
dust extinctions (Sect. 3.1), there are variations in the
spectrum of the Be component which show that intrinsic changes of the
Be envelope excitation
occur and possibly also influence the circumstellar extinction.

In Sect. 4  we discuss the formation regions of the various emission lines, 
and note that several strong absorption lines from He\,{\sc i}, O\,{\sc i}
and part of the Na\,{\sc i}\,D lines can also be attributed to the B0e star.
From the red Ca\,{\sc ii}\,(2) emission line triplet and various multiplets 
of Fe\,{\sc ii} emission lines  we have estimated  lower limits to the
disc radius of the `hidden' Herbig Be star. We also make estimates of
the outflow and accretion rates of the Be star.
 
In the discussion (Sect. 5) we compare our data of Z~CMa with those
of three Herbig stars with similar spectra: MWC 1080, V645 Cyg and V380 Ori. 
We also make some attempts to estimate the
evolutionary stage and mass of the B0e star from recent models for the
pre-main sequence evolution of massive stars. It appears possible to find 
models for the two components of Z~CMa which predict the
accretion rates (derived from the observations) and which also predict
a bipolar jet only for the B0e star.

\section{Observations}
\subsection{Photometry}
Since 1980 Z~CMa has been monitored closely in the Johnson $UBVR$ 
system by the ROTOR photometric monitoring program.  The 
observations in this program were made with three different 0.6 m 
telescopes at Mt. Maidanak Observatory, Uzbekistan.  During these 
observations, mostly made through a 15\arcsec\ circular diaphragm, 
the telescope was equipped with a single-channel photometer.
Standards and extinction reference stars were observed each night.  
The data were reduced with standard techniques at the Tashkent 
Astronomical Institute.  A detailed description of the ROTOR 
program and the employed measuring procedures can be found in 
Shevchenko (1989).

The ROTOR data on Z~CMa were supplemented with optical and infrared 
measurements on various photometric systems from the literature.  
When possible, all data were transformed to the ``standard'' Johnson/Cousins 
$UBVR_CI_C$ and near-infrared $JHKLM$ systems using the transformation 
formulae from Fernie (1983), Brand \& Wouterloot (1988), and 
Carpenter (2001).  The results are given and discussed in Sect. 3.1.

\subsection{Spectroscopy}
\begin{table*}
\centering
\caption{Log of spectroscopic observations of Z~CMa. We also list the $V$-band 
photometric measurement closest to the date of each spectroscopic observation.}
\small
\begin{tabular}{@{}ccccccc}
\hline
Telescope  & Range [\AA] &  Date & JD+2400000 & $R$ [m\AA] & $t_{\rm exp}$ [min] & $V$ [mag]\\
\hline 
SAO 6 m       & 4190--5490   & 22 Nov. 1991 & 48582.83  &   300 & -- & ~9.7\\
SAO 6 m       & 5960--7150   & 23 Nov. 1991 & 48583.83  &   300 & -- & ~9.7\\
ESO NTT       & 5720--9930   & 20 Dec. 1991 & 48610.838 & 279.5 & 20 & ~9.8\\
ESO NTT       & 5720--10030  & 21 Dec. 1991 & 48611.710 & 279.5 & 45 & ~9.8\\
ESO 1.5 m     & 4200--7800   & ~6 Dec. 1992 & 48962.823 &  1076 & ~4 & 10.3\\
ESO 1.5 m     & 6200--8700   & 10 Dec. 1992 & 48966.651 &  ~747 & ~8 & 10.3\\
ESO CAT       & 5860--5910   & 15 Dec. 1994 & 49701.355 & 107.0 & 45 & 10.2\\
ESO CAT       & 6521--6600   & 12 Dec. 1996 & 50429.151 & ~65.6 & 45 & 10.2\\
ESO CAT       & 5848--5918   & 13 Dec. 1996 & 50430.210 & ~58.8 & 45 & 10.2\\
ESO CAT       & 5848--5918   & 14 Dec. 1996 & 50431.235 & ~58.8 & 45 & 10.2\\
ESO CAT       & 3913--3955   & 14 Dec. 1996 & 50431.323 & ~39.3 & 45 & 10.2\\
ESO CAT       & 8467--8573   & 15 Dec. 1996 & 50432.235 & ~85.2 & 45 & 10.2\\
WHT           & 5220--9110   & 26 Dec. 1996 & 50443.725 & 266.7 & 20 & 10.3\\
WHT           & 5220--9110   & 26 Dec. 1996 & 50443.740 & 266.7 & 20 & 10.3\\
Tautenburg 2 m& 5600--10050  & 23 Jan. 2000 & 51567.422 & 230.0 & 20 & ~9.1\\
WHT           & 3640--5470   & 18 Feb. 2000 & 51593.361 & 177.2 & 30 & ~9.2\\
WHT           & 4100--8350   & 18 Feb. 2000 & 51593.423 & 232.0 & 30 & ~9.2\\
WHT           & 4600--9960   & 18 Feb. 2000 & 51593.447 & 278.4 & 30 & ~9.2\\
WHT           & 5330--11020  & 18 Feb. 2000 & 51593.498 & 325.8 & 30 & ~9.2\\
NASA IRTF     & 19000--41400 & 18 Dec. 2001 & 52262.139 & ~32.0 & ~5 & 10.0\\
ESO 3.6 m     & 6279--6321   & 31 Mar. 2002 & 52365.084 & ~42.0 & 30 & 10.0\\
\hline
\end{tabular}
%\noindent
%\flushleft
\end{table*}
In this paper we present the spectra of Z~CMa obtained with a range of 
telescopes:

(a)  Two low resolution spectra, taken on 9 and 10 December 1992 with 
the Boller and Chivens spectrograph on the 1.5 m telescope at ESO's 
La Silla Observatory. These spectra show emission in H$\alpha$, 
[O\,{\sc i}] 6300\,\AA,O\,{\sc i} 8446\,\AA\ and Ca\,{\sc ii}\,(2) 
(8500, 8545 and 8662\,\AA). The emissions of H$\beta$, P14
and Fe\,{\sc ii}\,(42), which are clear in the low resolution spectrum 
of the B5 star LkH$\alpha$ 220 (Tjin A Djie et al. 2001 (hereafter paper II), 
Fig. 5) are much weaker in this spectrum of Z~CMa. In contrast, the 
emission lines in the Z~CMa spectra of Whitney et al. (1993, Fig. 1). of 
Nov. 27, 1991 are stronger than those of LkH$\alpha$ 220.

(b) A high resolution blue spectrum (4190--5490\,\AA) obtained with the
Main Spectrograph on
the 6 m BTA  telescope of the SAO (Special Astrophysical Observatory in
Nizhnii-Arkhyz) on
22 Nov. 1991 and a high resolution H$\alpha$ profile obtained with the same
instrument on 23 Nov. 1991. Similar to the spectrum of Whitney et
al. (1993), which was taken a few days later, the SAO spectrum shows 
strong emission components in H$\alpha$, H$\beta$, H$\gamma$ and
in several multiplets of Fe\,{\sc ii}. These spectra are supplemented by high
resolution red {\it ESO Multiple-Mode Instrument} (EMMI) spectra, obtained 
on 20 and 21 Dec. 1991 with the 3.5 m ESO {\it New Technology Telescope} (NTT) 
and extracted by us from the ESO science archive.

(c)  High resolution echelle spectra have been secured with
 the {\it Coud\'e echelle
spectrograph} (CES) mounted on the 1.4 m {\it Coud\'e Auxiliary Telescope} 
(CAT) at La Silla:
 (1) Na\,{\sc i}\,D absorption profiles of Dec. 15, 1994 and of Dec. 13 
and 14, 1996
 (2) An H$\alpha$ emission profile of Dec. 12, 1996
 (3) A Ca\,{\sc ii}\,K emission and absorption profile of Dec. 14 1996
 (4) Ca\,{\sc ii} (8498\,\AA) and Ca\,{\sc ii} (8542\,\AA) (blended with resp. P16 and P15)
 emission profiles of Dec. 15, 1996.
 Additionally, an emission profile of the [O\,{\sc i}] 6300\,\AA\ line has been 
obtained with the same instrument connected to the ESO 3.6 m telescope
on March 31, 2002.

(d)  A high resolution red spectrum (5470--9110\,\AA) taken on Dec. 26,
 1996 (close to the dates of the observations mentioned under c) with the 
{\it Utrecht Echelle Spectrograph} (UES) on the 4.2 m WHT at the Northern 
Hemisphere Observatory on La Palma was kindly made available
to us by Drs. H. van Winckel and G. Meeus of the Katholieke Universiteit 
Leuven (Belgium).

(e)  A high resolution red spectrum (5600--10050\,\AA) was obtained
on January 23, 2000 with the Coud\'e Echelle Spectrograph of the 2 m 
Alfred Jensch telescope of the Th\"uringer Landessternwarte `Karl Schwarzschild'  
in Tautenburg, Th\"uringen. This spectrum was taken during a very bright phase 
of Z~CMa ($V$ $\sim$ 9.1\mag), comparable to the outburst of Feb. 1987 
($V$ $\sim$ 8.7\mag).  The spectrum covers the wavelength range from 
about 560 to 1005 nm.  Using a Tek CCD Chip with 1024 $\times$ 1024 pixels 
and a slit-width of two arcseconds, the two-pixel resolution of the spectrum 
is about $\lambda$/($\delta \lambda$) = 35,000.  Standard IRAF routines were 
used in order to extract and wavelength calibrate the spectrum.

(f)  Four high resolution echelle spectra, centered at 4300\,\AA, 5200\,\AA, 
6200\,\AA\ and 8200\,\AA\ were obtained on Feb. 18, 2000 with the UES 
on the 4.2 m WHT.  A number of line profiles
from these spectra will be compared with the corresponding
profiles from the spectra mentioned under (b)--(e) and with
those published in the literature. Apart from the blue range, these
spectra extend in the red up to 11020\,\AA. However, some lines (e.g.
[O\,{\sc i}] 6300\,\AA) could not be observed
in these spectra because there is no wavelength overlap of the
successive orders.

(g)  A near-infrared (1.9--4.1 $\mu$m) spectrum has been taken with 
the SpeX spectrograph (Rayner et al. 2003) attached to the 
3.0 m NASA {\it Infrared Telescope Facility} (IRTF) 
of the Mauna Kea Observatory on Dec. 18, 2001. In this spectrum
we observe Br $\alpha$, Br $\gamma$, Pf $\gamma$ and several 
CO lines in emission.

Observational details of all spectroscopic observations are reviewed 
in Table~1.

\section{Photometry and Interpretation}
\subsection{Variability and Lightcurve}
\begin{figure}
\vspace*{0.15cm}
\centerline{\psfig{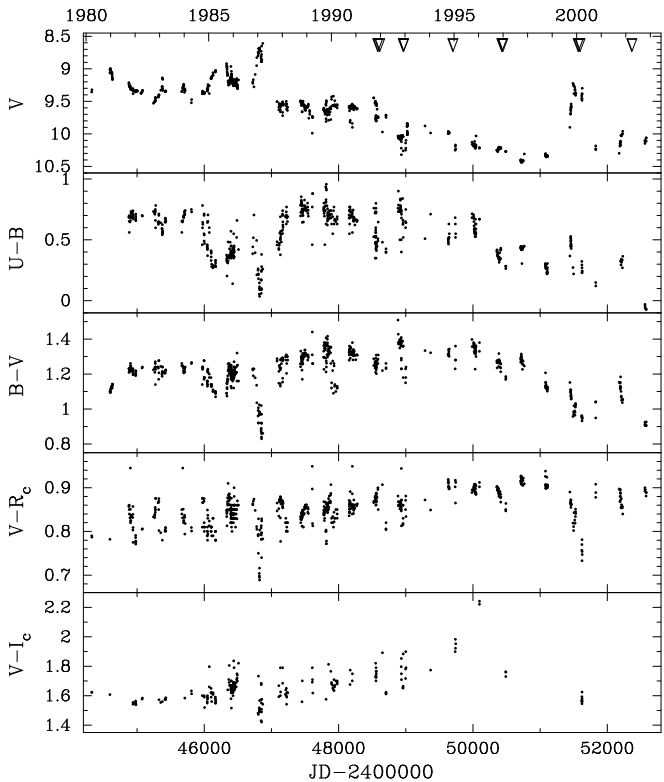}}
\caption[]{Optical ($UBVR_cI_c$) light-curve of Z~CMa spanning the 
period 1980--2003.  Typical errors in magnitude and colours are 
of the order of 0.02 mag.  The triangles in the top panel indicate 
the dates of the spectroscopic observations listed in Table~1.}
\end{figure}
Figures 1 and 2 show the optical ($UBV(RI)_C$) and near-infrared ($JHK$) 
lightcurve of Z~CMa between 1980 and 2003.  For a detailed description 
of the photometric behaviour of Z~CMa before the 1987 outburst, we 
refer the reader to the papers of Covino et al. (1984), Hessman et al. (1991), 
and Lamzin et al. (1998).  From Fig. 1 we conclude that whereas 
before 1987 the system brightness appears 
relatively constant, from Feb. 1987 to Dec. 2000 the brightness of the 
star is steadily declining by about 0.1\mag per year.  Clear variability 
is seen superimposed on this trend, such as a local minimum around 
December 1992.  There appears to be a strong correlation between the 
observed brightness and colour variations in Z~CMa.  We illustrate 
this point in Fig.~3, where we give the $V$ vs. $(U-B)$, $(B-V)$, 
$(V-R_C)$ and $(V-I_C)$ distributions for the time-interval 1980--2003.  
Although some correlation is present in these diagrams, the spread 
of the data-points is much larger than expected from the observational 
errors ($\sim$ 0.02\mag).
\begin{figure}
\vspace*{0.15cm}
\centerline{\psfig{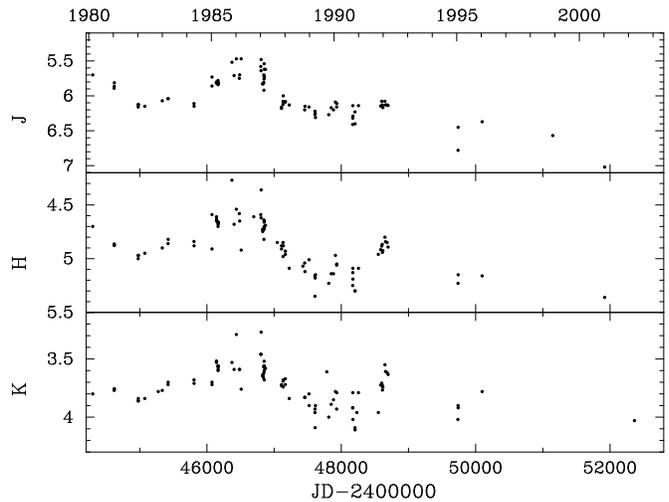}}
\caption[]{Near-Infrared ($JHK$) light-curve of Z~CMa spanning the 
period 1980--2003.  Typical errors in these data are about 0.05 mag.}
\end{figure}

A closer inspection shows that between Feb. 1987 and Nov. 1991 the
photometric variations closely follow a behaviour of redder colours 
when the system gets fainter, but that the data of Dec. 1999 and 
March 2001 are shifted to somewhat lower values of $(B-V)$ and $(U-B)$ 
than defined by the reddening line.  We infer that either variable 
circumstellar extinction or a slow cooling-down of the FUor component 
in Z~CMa may explain the long-term overall trend of the photometric 
variation between Feb. 1987 and Dec. 2000.  However, on a shorter 
time-scale of weeks and days the variations seem erratic and hard to
explain. This may not be surprising, since we probably are observing
the result of two independently varying sources, a FUor secondary 
component which usually dominates the visual continuum and a Be type 
primary star in a large dust cocoon, which is dominating the
near and far infrared part of the spectrum (Koresko et al. 1991).
Fig. 2, displaying the variation of the near-infrared magnitudes, shows 
that in 1991 the $H$ and $K$ band fluxes reach a local maximum,
which is more pronounced than that in the $V$ band. This suggests that
the circumstellar extinction of the cocoon around the primary component 
has decreased so far, that the visual flux from that component could 
increase its contribution to the total flux in
the visual somewhat.
\begin{figure*}
\vspace*{0.15cm}
\centerline{\psfig{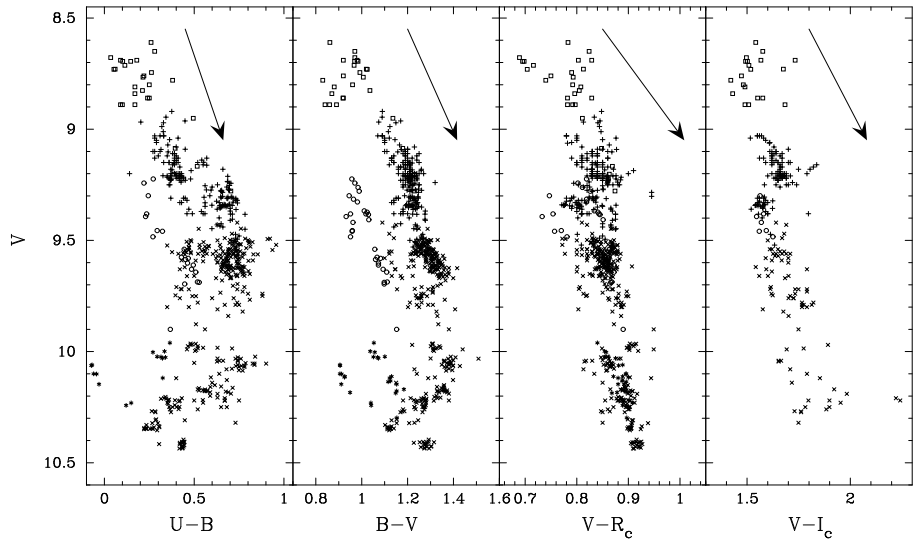}}
\caption[]{Optical ($UBVR_cI_c$) colour-magnitude diagram of Z~CMa 
using photometric data from the period 1980--2003. Data from 
Jan. 1980--Jan. 1987 are indicated by the plusses, data from 
the Feb. 1987 outburst by open squares, data from 
Mar. 1987--Oct. 1999 by the crosses, data from the Nov./Dec. 1999 
outburst by the open circles, and data points obtained after 
Feb. 2000 are indicated by the asterisk symbols.  The 
arrows indicate the direction of interstellar reddening}
\end{figure*}

Apart from the long-term variations discussed above, the dominant features 
in the optical light-curve of Z~CMa are the large and fast rises 
in visual flux in Feb. 1987 and Jan. 2000 (Fig.~1). 
The latter maximum started with a fast rise, first observed in 
Oct. 1999, over 0.7\mag\ in $V$ taking 46 days.   The star 
remained at this high brightness for at least 180 days.  When 
extrapolating to earlier dates, the total increase in $V$-brightness 
is $\sim$1.1\mag, but the time-scale of this larger variation has not been 
observed. The time-scale of the rise in brightness over 0.7\mag\ during 
Jan./Feb. 1987 is also not known, but the duration of the higher 
brightness phase was also about 6 months (Hessman et al. 1991, Fig.  1).

We note that the large and fast rises in the visual fluxes in 
Feb. 1987 and Jan. 2000, do not originate in a large, fast decrease in 
circumstellar extinction; it will be shown by the spectroscopy (Sect. 4)
that they are related to intrinsic changes of the primary Be star
component. 

\subsection{The SEDs in the visual}
From the spectral polarization data measured in Nov. 1991 by Whitney et
al. (1993), the authors
derived that at this date 75\% of the visual flux was contributed by
the FUor secondary source and
that the remaining 25\% was due to the Be-type primary star. By
assuming that the fast rise in visual
brightness of Feb. 1987 (Hessman et al. 1991) was entirely due to a
rise in the primary brightness
only, they derived that on Feb. 19, 1987 the contribution of the
primary Be star to the total visual
flux was 66\%. In the same way we derive from the fast rise in
Nov./Dec. 1999 that at maximum brightness
in Jan. 2000 ($V$ $\sim$ 9.1\mag) the contribution of the primary Be star to the
total visual flux was $\sim$ 60\%.
\begin{table}
\centering
\caption{Radii and luminosities of the B0IIIe star on 20 Feb. 1987, 23 Jan. 2000, 
and 22 Nov. 1991 for $R_{\rm cs}$ = 3.1, $R_{\rm cs}$ = 4.2 and $R_{\rm cs}$ = 6.0.  
The uncertainties in the computed values of $R/R_\odot$ and $\log (L/L_\odot)$, 
dominated by the 14\% error in the distance to CMa~R1 (paper I), are less than 
10\% and 3%, respectively.  Errors in the derived values of $T_{\rm eff} are 
around a few thousand~K.}
\tabcolsep0.11cm
\small
\begin{tabular}{@{}ccccccc}
\hline
Obs. Date  & $V_{\rm obs}$ [\mag] & $R_{\rm cs}$ & $V_0$ [\mag] & $R/R_\odot$ & $\log (L/L_\odot)$ & $T_{\rm eff}$ [K]\\
\hline 
02/20/87 &  8.8  &  3.1  &  5.39  &  12.3  &  4.81  & 26,000\\
01/23/00 &  9.2  &  3.1  &  5.70  &  10.6  &  4.68  & 26,000\\
11/22/91 &  9.8  &  3.1  &  6.14  &  ~8.0  &  4.57  & 28,000\\
\\
02/20/87 &  8.8  &  4.2  &  4.50  &  18.5  &  5.16  & 26,000\\
01/23/00 &  9.2  &  4.2  &  4.72  &  16.7  &  5.07  & 26,000\\
11/22/91 &  9.8  &  4.2  &  4.80  &  16.1  &  5.04  & 28,000\\
\\
02/20/87 &  8.8  &  6.0  &  3.03  &  33.6  &  5.81  & 28,000\\
01/23/00 &  9.2  &  6.0  &  3.12  &  32.4  &  5.78  & 28,000\\
11/22/91 &  9.8  &  6.0  &  2.61  &  37.5  &  6.01  & 30,000\\

\hline
\end{tabular}
%\noindent
%\flushleft
\end{table}

In the UV the spectrum is dominated by the FUor companion. Earlier
obtained UV spectra could be matched with a model of an accretion disc 
with a late F-type spectrum (Hartmann et al. 1989; Kenyon et al. 1989). 
We have made a similar fit of two IUE spectra (taken on 17 April 1979 when
$V$ $\sim$ 9.3\mag\  and 19 March 1988 when $V$ $\sim$ 9.6\mag). 
Good fits were obtained with a F5Ib star, an 0.02 R$_\star$ thick boundary 
layer with $T_{\rm eff}$ = 10,500 K and an optically thick disc with 5400 K maximum
temperature. We assumed a foreground excess of 0.16\mag\ with $R_{\rm is}$ = 3.1
and a foreground CMa~R1 cluster excess of 0.20\mag with 
$R_{\rm cs}$ = 4.2 (Shevchenko et al. 1999; Paper I). The accretion 
rate for the FUor is then $\sim$ 3 $\times$ 10$^{-5}$ M$_\odot$~yr$^{-1}$.

From the observed $UBVRI$ photometry, the above mentioned
foreground excess and the relative flux contributions of a B0III primary and 
a FUor (F5Ib) secondary component we have calculated the extinction-free 
$UBVRI$ fluxes of both components in 1987, 1991 and 2000 for various values 
of  $R_{\rm cs}$ (3.1, 4.2 and 6.0). We assumed here intrinsic 
colours $(B-V)_0$ = $-$0.29\mag\ for a B0III star and $(B-V)_0$ = 0.37\mag\ 
for the FUor (Schmidt-Kaler 1982). 
The extinction-free $UBVRI$ distributions of the B-type primary star can
be matched with the Kurucz model flux ratios of a late O or early B-type 
star, which gives us an effective temperature, a radius and a
luminosity for each of the dates of spectroscopic observations (Table 2). 
We will use these results in Sect. 4 and in the discussion on the evolutionary 
status of Z~CMa (Sect. 5).

     In reality the values of $R/R_\odot$ and $\log (L/L_\odot)$ of 
Table 2 may be somewhat higher because so far
we have not accounted for the fact that we only receive the light from
the star after it has been
scattered and polarized by the dust walls of the cavity in the cocoon.
The correction for this
reduction is uncertain, but because of the upper limit of 10\% for the
degree of polarization (Whitney et al. 1993) we
estimate that the scattering angle will not be larger than 20 degrees,
so that the flux reduction
due to scattering will not be more than 5\% if the scattering grains
are more or less aligned
(see Figs. 8 and 9 of the paper of Voshchinnikov \& Farafonov 1993).
Our estimate that the scattering angle is small is supported by the
relatively large radial velocities of the [N\,{\sc ii}] and 
[S\,{\sc ii}] emission lines in the bipolar outflow along the axis
of the cavity (Poetzel et al. 1989).

\subsection{The mass loss of the Be primary component}
   From Jan. 1981 Z~CMa photometry on the Walraven $WULBV$ 
system (de Winter et al. 2002) we can make an estimate of the mass 
loss from the early Be-type companion. After transformation of the 
flux in the Walraven $V$-band to the Johnson system we find that
the visual magnitude in this period was 9.0\mag, which suggests that the continuum 
of the B star dominated the visual part of the spectrum at that date. 
After correction for foreground extinction, the Balmer jump $D_B$,
derived from the Walraven photometry is 0.16\mag. Predicted values of
the Balmer jump based on model calculations by Mihalas (1966) can be 
found in Fig. 2 of the paper of  Garrison (1978).
For B0 and $\log g$ = 3 the theoretical value of the Balmer jump is close
to 0.15\mag, but for $\log g$ = 4 the predicted value is already 
0.29\mag, which implies that $\Delta D_B$ for B0III has an upper
limit of 0.13\mag. We estimate the mass loss with the method of Garrison
and assume a temperature of 10,000 K for the outflow region. Then the 
upper limit of $\Delta D_B$ leads to an upper limit of
1.6 $\times$  10$^{37}$ cm$^{-6}$ for the volume emission measure and with an
outflow velocity of 1000 km~s$^{-1}$ in the H$\alpha$ emission region 
(see Sect. 4) and a radius of 9.4 R$_\odot$ we have an upper limit of
1 $\times$ 10$^{-4}$ M$_\odot$~yr$^{-1}$ for the mass loss in the wind. 
The mass loss is proportional with the outflow
velocity in the emission region, so that the actual upper limit can
easily be twice as low.

\section{Analysis of the Spectra}
\subsection{Results}
\begin{figure}
\vspace*{0.15cm}
\centerline{\psfig{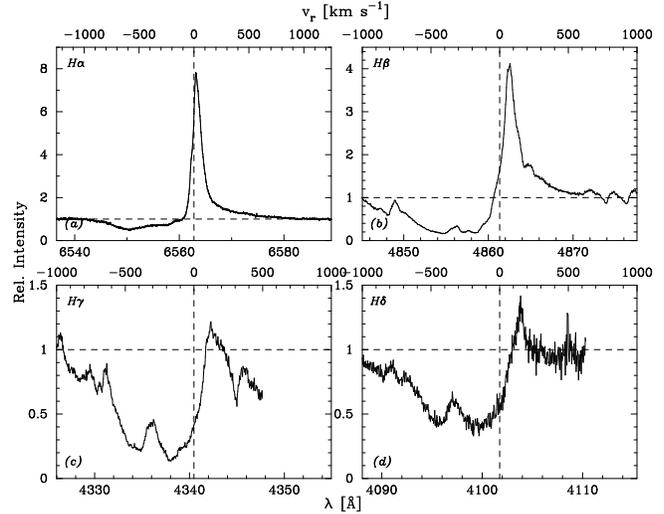}}
\caption[]{H\,{\sc i} Balmer line profiles of  Z~CMa. (a) H$\alpha$ 
line profile obtained on 12 Dec. 1996, (b) H$\beta$ profile obtained 
on 18 Feb. 2000, (c) H$\gamma$ profile obtained on 18 Feb. 2000, 
(d) H$\delta$ profile obtained on 18 Feb. 2000.  The dashed 
horizontal and vertical lines indicate the continuum level, and 
the rest-wavelength of the observed line, respectively.}
\end{figure}
\begin{figure}
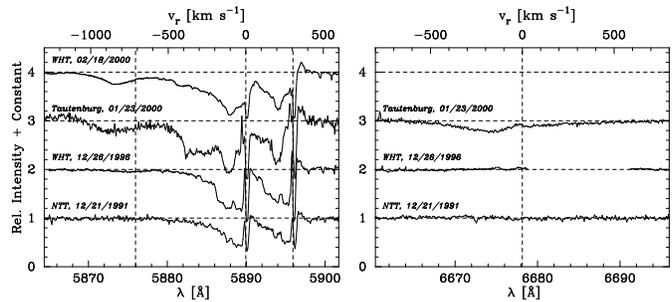

\vspace*{0.15cm}
\centerline{\psfig{figure=zcma_fig5a.ps,height=3.9cm,angle=270}
            \hspace*{0.05cm}
            \psfig{figure=zcma_fig5b.ps,height=3.9cm,angle=270}}
\caption[]{Na\,{\sc i} and He\,{\sc i} line profile variations 
in Z~CMa between 1991 and 2000.  The left panel shows the 
5889.95 and 5895.92\,\AA\ Na\,{\sc i}\,D lines, together with 
the 5875.97\,\AA\ He\,{\sc i} line, whereas the right panel 
shows the spectral region around the 6678.15\,\AA\ He\,{\sc i} 
line.  Strong He\,{\sc i} absorption is only present in the 
Jan. and Feb. 2000 spectra.}
\end{figure}
\begin{figure}
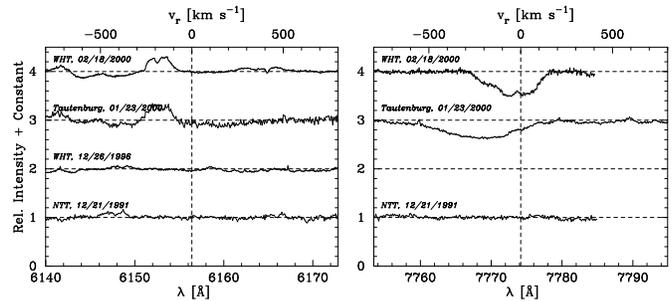

\vspace*{0.15cm}
\centerline{\psfig{figure=zcma_fig6a.ps,height=3.9cm,angle=270}
            \hspace*{0.05cm}
            \psfig{figure=zcma_fig6b.ps,height=3.9cm,angle=270}}
\caption[]{O\,{\sc i} line profile variations in Z~CMa between 
1991 and 2000.  The left panel shows the observed variability 
in the O\,{\sc i} 6156.41\,\AA\ line, whereas the right-hand 
panel illustrates the behaviour of the O\,{\sc i} 7774.17\,\AA\ 
line.}
\end{figure}
Table~4 gives the data of the lines in the red spectra
(longward 5650\,\AA) in the Z~CMa
spectra for $V$ $\sim$ 9.1\mag, for $V$ $\sim$ 9.7\mag\ and 
for $V$ $\sim$ 10.3\mag. The tables 
contain equivalent widths (EWs), half-widths 
(FWHM) and P Cygni outflow velocities of the lines. In Tables 3 the
corresponding data of the lines in the blue spectra are listed. 

In order to learn more about the structure of the envelope of the
early Be type primary we
discuss the various lines and their profiles in the high resolution
red and blue spectra of Jan.
and Feb. 2002 ($V$ $\sim$ 9.1\mag) and then compare the results with the
corresponding red and blue
spectra, observed in Nov. (blue) and Dec. (red) 1991 ($V$ $\sim$ 9.6\mag), and
with the profiles of several
characteristic lines, secured in Jan. 1996 ($V$ $\sim$ 10.2\mag).

\subsection{The outburst spectra of  2000 and 1987}
\subsubsection{The emission lines}
    In discussing the lines we first should know of which stellar
component they originate.  The presence of N\,{\sc i} (8629 and 8683\,\AA) 
emission is an indication for a hot atmosphere, since
the lines have been only observed for (early) B stars (Hamann \&
Persson 1992b). Similarly strong, high excitation Fe\,{\sc ii} 9997.57\,\AA\  
emission has only be observed in spectra of
early Be type stars, such as $\gamma$ Cas (B0IV) (Viotti et al. 1998).
This line is probably formed by fluorescence due to pumping by Ly$\alpha$ 
(see e.g. Rodr\'{\i}guez-Ardila et al. 2002).  Whitney et al. (1993) and 
Garcia et al. (1999) have shown that the emission lines are associated 
with the primary Be-star. The main part of this emission will
come from the extended envelope of this star, but the forbidden lines 
of [N\,{\sc ii}], [S\,{\sc ii}] and [O\,{\sc i}] have been shown to be 
formed in the bipolar outflow (or a jet) from the primary (Poetzel et al. 1989; 
Garcia et al. 1999). From spectral imaging Bailey (1998) revealed that 
also a substantial part of
the central region of the H$\alpha$ emission is emitted by the jet. 
The peak of the line is redshifted by +82 km~s$^{-1}$ with respect to 
the laboratory wavelength. 
The wings of the line have their origin in the extended atmosphere
of the primary Be-star. Especially the red wing extends very far, up
to nearly 1000~km~s$^{-1}$. 
The emission peak of H$\beta$ shows a similar redshift of +75 km~s$^{-1}$. 

We assume that the other Balmer (and Paschen) lines will be formed in 
the same region as H$\alpha$.  A strong argument in favour of this
is that all Balmer lines (from H$\alpha$ to H15) in the WHT spectrum have 
a similar outflow velocity pattern.  From the last resolved Balmer 
lines ($n$ $\sim$ 35) we derive from the
Inglis-Teller relation that $n_e$ $\la$ 
2.4 $\times$ 10$^{11}$ cm$^{-3}$ in the region of their formation.
For the Paschen lines, which have strong emission components (lower
continuum in red part of the spectrum), $n$ = 23 which corresponds to  
$n_e$ $\approx$ 10$^{13}$ cm$^{-3}$.

  The emission line of O\,{\sc i} at 8446.5\,\AA\ is expected to be correlated with
the H$\alpha$-emission, because the $3d^3D^0$ state of O\,{\sc i}, which can be
pumped by Ly$\beta$ (Polidan \& Peters 1976), cascades down succesively by the
emission of 11291\,\AA\ to $3p^3P$, that of 8446\,\AA\ to $3s^3S^0$ 
and finally by that of 1302\,\AA\ to the $3P$ ground state. Its EW varies from
negligible at Apr. 1989 ($V$ $\sim$ 9.7\mag) to 1.6\,\AA\ in Dec. 1987 ($V$ $\sim$ 9.5\mag, 
Hamann \& Persson 1992b), 1.2 in Dec. 1991 
($V$ $\sim$ 9.6\mag), 1.5\,\AA\ in Jan. 1996 ($V$ $\sim$ 10.2\mag, Teodorani et al. 1997) 
to 1.4\,\AA\ on Jan 2000 ($V$ $\sim$ 9.1\mag).
For the last three dates, the EW of H$\alpha$ varied from 19.1, 32.5 to 42.2\,\AA\, 
which does not seem to confirm the expected correlation between the EWs 
of O\,{\sc i} and H$\alpha$ for Z~CMa.  It should be kept
in mind though that the H$\alpha$ emission peak contains an unknown
component, which is emitted
by the outflow region (Bailey 1998). The variation of the H$\alpha$ emission
EW is therefore partly due to variation in the outer atmosphere, whereas 
the emission of O\,{\sc i} (8446.5\,\AA) originates in
the inner part of the atmosphere. 

        Next to the Balmer emission lines, the Ca\,{\sc ii}\,(2) triplet lines
have by far the strongest emission fluxes in the spectra of Z~CMa. Although 
the three lines are blended by the Paschen lines
P16, P15 and P14, the contributions from these lines to the flux are
small (their EWs are about 5\% of the total EW). The lines have P-Cygni 
profiles with outflow velocities around 500 km~s$^{-1}$ and
half-widths of 107, 110 and 132 km~s$^{-1}$ for 8498, 8542 and 8662\,\AA\
respectively. The line shapes are very
asymmetrical: the red wings extend to more than 8\,\AA\ from the line
center, which suggests that the lines
are formed in a turbulent region.

      We confirm the presence of many emission lines from Fe\,{\sc ii} of
multiplets 27, 28, 32, 35, 36, 37, 38, 40, 42, 43, 46,
48, 49, 55, 73, 74 and 199, several lines of Cr\,{\sc ii} multiplet 30, 43 and 44
and a few lines of Ti\,{\sc ii}\,(20) and Sc\,{\sc ii}\,(29).
Apart from multiplet 73 in the red part of the spectrum,
most of these lines are also given in
the line list of the high state spectrum of Hessman et al. (1991). We
discuss the variation of the Fe\,{\sc ii} lines
in Sect. 4.6. In the red part of the spectrum we also observed fairly
strong emission of the lines of Fe\,{\sc i} multiplet 60.

\subsubsection{Source of the broad absorption troughs of  He\,{\sc i} and O\,{\sc i}}
The lines of He\,{\sc i} (7065, 6678 and 5876\,\AA) (Fig. 5) 
and O\,{\sc i} (7773 and 6156\,\AA) (Fig. 6) have relatively broad
absorption features.  They also show a weak emission at a small
redshift (+65 km~s$^{-1}$) from the
laboratory wavelength and a broad ($-$450 km~s$^{-1}$) 
blueshifted absorption trough and therefore have
the shape of P Cygni profiles. There are several reasons to believe
that these profiles are formed in the envelope of the primary Herbig B0e star: 
In Fig. 10 of the paper by Hessman et al. (1991)
we see that in the post-eruption phase the 6678\,\AA\ He\,{\sc i} singlet line has
disappeared and replaced by
two Fe\,{\sc i} lines with the double structure which is characteristic for
the FUor component. The broad
profile for the high state (in the same Fig. 10) shows only weak
indications of these double lines, which suggests that during this
state the continuum of the primary star is higher than that of the
secondary
FUor component (see also Sect. 3.2), so that the absorption line
contribution from the primary (Be) component is dominating. In fact it
was estimated by Whitney (1993) that in the high state the
contribution of the FUor component to the visual continuum is only one
third of the observed
continuum and therefore one half of the contribution by the primary Be
star.

   The 7065\,\AA\ and 5876\,\AA\ He\,{\sc i} triplet lines in the Tautenburg spectrum
show the same profiles
as that of the 6678\,\AA\ He\,{\sc i} singlet line. Although the 6678\,\AA\ line is
outside the orders of the
WHT spectrum, we can see from the comparison of the Tautenburg and WHT
spectra
that the profile of the 5876\,\AA\ line is the same in both spectra, so
that the He\,{\sc i} line did not
change over a period of one month, in contrast to the neighbouring Na\,{\sc i}
profiles.
Because of its high excitation energy (23 eV), the emission
components in the He\,{\sc i} lines will originate from the hot neighbourhood
of the stellar surface. The broad blue-shifted absorption components
must be interpreted as a strong and fast outflow with a broad (perhaps
turbulent) velocity spectrum.
We also note here that no line of He\,{\sc ii} (e.g. 4686\,\AA) was 
found in our optical spectra so that there are no direct indications for 
the presence of a chromosphere or transition region.

 The O\,{\sc i} (7773\,\AA) triplet has been observed in absorption for several
Herbig Ae/Be stars (Felenbok et al. 1988; Hamann \& Persson 1992b). In Z~CMa, 
this line shows indications of emission
components, redshifted by $\sim$ 70 km~s$^{-1}$ and the absorption components are
very similar to that of the 6678\,\AA\ He\,{\sc i} line.  This indicates that the
line is not necessarily formed by the FUor component. It is also clear 
that in Jan. 2000 the continuum
of the primary B0e star was much higher than that of the FUor at the
time of the observation.

Hamann \& Persson (1992b) suggest that the O\,{\sc i} profile is formed in a
highly turbulent envelope region.
 This may also be concluded from spectral surveys over many stars with
different spectral and luminosity class (e.g. Slettebak 1986; Jaschek et al. 1993). 
These authors show that the only early
B-type stars with O\,{\sc i} (7773\,\AA) in absorption are the shell stars. 
In addition it has been shown by
Faraggiana et al. (1988) that for this line EWs of the order of 2\,\AA\ 
are only found for stars of spectral type earlier than B1. For Z~CMa the 
maximum velocity of the outflow is  $\sim$ 500 km~s$^{-1}$ in the Tautenburg 
spectrum. Because the excitation potential of the O\,{\sc i}
lines is only 10.7 eV, the inner border of the turbulent region were
the emission of O\,{\sc i} is formed will be further away from the stellar 
surface than that of He\,{\sc i} (6678\,\AA). The same profile as for
the O\,{\sc i} (7773\,\AA) triplet is found for the 6157\,\AA\ triplet of 
O\,{\sc i}, which, when in emission, cascades down by 7773\,\AA\
emission from the $3p^5P$ state to the metastable $3s^5S^0$ 
state at $\chi_i$ = 9.5 eV.

\subsubsection{The Na\,{\sc i}\,D, Ca\,{\sc ii}\,K and K\,{\sc i} lines}
The observed Na\,{\sc i}\,D profiles (Fig.~5) consist of at least three 
components: a narrow absorption component, an emission component and a
broad absorption
component, composed of several blue-shifted lines. The narrow
components of D$_1$ and D$_2$
are shifted by $\sim$ 27 km~s$^{-1}$, which is close to the radial velocity of
Z~CMa. We assume that they
are due to absorption by the foreground interstellar medium and the local CMa
cloud. Their FWHM is $\sim$ 24 km~s$^{-1}$. The EWs of the narrow components 
(Table 4) can be used to estimate the Na\,{\sc i}
foreground column densities with the help of the Str\"omgren (1948)
doublet method. By assuming  a galactic abundance ratio $N$(H)/$N$(Na) 
of 5 $\times$ 10$^{5}$ we can derive an average
interstellar gas column and, with the empirical relation of Hobbs
(1974), we can obtain a value
for $E(B-V)$ of 0.32, which is in line with the foreground excesses of
the other stars in CMa~R1 (papers I and II).

In the high resolution spectra obtained at ESO ($V$ $\sim$ 10.2\mag) and
Tautenburg ($V$ $\sim$ 9.1\mag), the emission components of the 
Na\,{\sc i}\,D lines are narrow and on both
sides of the narrow absorption components.  We suspect that they are
the emission
components (intersected by the interstellar absorption lines) of Na\,{\sc i}\,D P-Cygni
profiles. 

The deepest and narrowest absorption component extends from $-$30 to $-$200 km~s$^{-1}$ 
with respect to the interstellar component with a  FWHM of 65 km~s$^{-1}$. 
According to Whitney et al., the deep absorption component of the D-lines 
is associated with the secondary (FUor) component. It
causes a strong increase in polarization of the combined spectrum over
this wavelength interval
because the deep Na\,{\sc i}\,D absorption lines  of the secondary source cannot
dilute the polarized
continuum emission of the primary. The deep absorption components
around 5896 and 5886\,\AA\ 
($F/F_c$ $\approx$ 0.2--0.3) will therefore almost certainly be formed around the
secondary star. This is very
similar to the Na\,{\sc i}\,D line profiles of the FUors FU Ori and V1057 Cyg
(see e.g. Welty et al. 1992,
Figs. 6 and 7), where the maximum outflow velocities of the absorption
trough are $-$200 and $-$250 km~s$^{-1}$ respectively. The (P Cyg) emission peaks are
also usually seen for the Na\,{\sc i}\,D profiles of FUors.
At bright phases of Z~CMa (e.g. the `high state') the blue absorption
wings (shortward 5886\,\AA\ and 5890\,\AA), however, may receive significant 
contributions from the primary source, because then the continuum
of the secondary star is lower than that of the primary. In Jan. and
Feb. 2000 the absorption wings extend from $-$200 to $-$500 km~s$^{-1}$.  
The fact that this new wing shows up simultaneously with the maximum 
in brightness of the primary star suggests that this outflow component 
occurs in the outer atmosphere of the B0 III star.  The strength of this 
wing varies within one month (Fig. 5).  

The K\,{\sc i} (7698.96\,\AA) line profile (not shown here) shows the opposite 
behaviour.  In Jan. 2000 it consists of a (broad) emission component and a 
narrow one (FWHM $\approx$ 12 km~s$^{-1}$). If the absorption component is 
interstellar, its EW leads to a K\,{\sc i} abundance which is 56 times lower 
than the Na\,{\sc i} abundance derived here. This is 2.8 times 
lower than the solar abundance of  K\,{\sc i}. In the WHT spectra of Feb.
2000 a second, broader (FWHM $\approx$ 33 km~s$^{-1}$) absorption component has
appeared, which is blue-shifted by about $-$33 km~s$^{-1}$ with respect to the
first component. This broad component, which could be formed in the outer 
disc or circumstellar region is also present in the spectra of 1991 and 1996.
It demonstrates again that during the outburst  variability in the
outer shell of the primary Be star occurs within a time-scale of less
than one month.  The Ca\,{\sc ii}\,K profile of the Feb. 2000 WHT spectrum 
(Fig.~8) shows an outflow velocity pattern which is similar to 
that of the Na\,{\sc i}\,D lines in the same spectrum, but completely different
from those of the Fe\,{\sc ii}\,(42) and the Balmer lines. This suggests that 
the Ca\,{\sc ii}\,K line is formed over the same outer
envelope or disc region as Na\,{\sc i}\,D, which is in line with the 
comparable excitation potentials of the two lines.
\begin{table*}
\centering
\caption{Lines detected in the blue spectra of Z~CMa (bl = blend). Lines included in square 
brackets indicate forbidden transitions.}
\small
\begin{tabular}{@{}clccccccccccccc}
\hline
$\lambda$ [\AA] & Ident. &  \multicolumn{3}{c}{WHT, 01/23/2000}   && \multicolumn{3}{c}{BTA, 11/22/1991} &&  \multicolumn{3}{c}{WHT, 12/26/1996}\\
\noalign{\smallskip}
\cline{3-5} \cline{7-9} \cline{11-13}
\noalign{\smallskip} 
           &  & EW & FWHM & $v_{\rm out}$ & \, & EW & FWHM & $v_{\rm out}$ & \, & EW & FWHM & $v_{\rm out}$\\
           &  & [\AA] & [km~s$^{-1}$] & [km~s$^{-1}$] & \, & [\AA] & [km~s$^{-1}$] & [km~s$^{-1}$] & \, & [\AA] & [km~s$^{-1}$] & [km~s$^{-1}$]\\
\hline
 3933.66  &  Ca\,{\sc ii}\,(1)      & $-$0.96  & ~97 & $-$600 & \, &          &      &        & \, &   $-$1.64$^{\dagger}$ & ~71 &  $-$650\\
 4233.17  &  Fe\,{\sc ii}\,(27)     & $-$0.44  & ~80 & $-$106 & \, &  $-$0.31 &  130 &        & \, &           &     &	\\
 4278.13  &  Fe\,{\sc ii}\,(32)     & $-$0.19  & ~88 &        & \, &  $-$0.17 &  ~68 &        & \, &           &     &	\\
 4290.22  &  Ti\,{\sc ii}\,(41)     & $-$0.28  & ~53 & $-$107 & \, &  $-$0.08 &  ~67 &        & \, &           &     &	\\
 4295.17  &  Fe\,{\sc i}\,(41) +    & $-$0.28  & ~53 & $-$107 & \, &  $-$0.07 &  ~49 &        & \, &           &     &	\\
 4295.90  &  Ti\,{\sc ii}\,(20)\,(bl) &          &     &        & \, &          &      &        & \, &           &     &	\\
 4306.67  &  Fe\,{\sc i}\,(42) +    & $-$0.24  & ~44 &  $-$98 & \, &  $-$0.08 &  ~98 &        & \, &           &     &	\\
 4307.15  &  Ti\,{\sc ii}\,(41)\,(bl) &          &     &        & \, &          &      &        & \, &           &     &	\\
 4320.54  &  Ti\,{\sc ii}\,(41) +   & $-$0.22  & ~67 &  $-$79 & \, &  $-$0.07 &  ~62 &        & \, &           &     &	\\
 4321.95  &  Sc\,{\sc ii}\,(15)\,(bl) &          &     &        & \, &          &      &        & \, &           &     &	\\
 4341.77  &  H$\gamma$              & $-$1.68  & 158 & $-$704 & \, &  $-$0.48 &  115 & $-$760 & \, &           &     &	\\
 4351.75  &  Fe\,{\sc ii}\,(27)     & $-$0.56  & ~79 &  $-$95 & \, &  $-$0.18 &  115 &        & \, &           &     &	\\
 4358.36  &  [Fe\,{\sc ii}]         & $-$0.04  & ~53 &        & \, &  $-$0.03 &      &        & \, &           &     &	\\
 4368.25  &  Fe\,{\sc i}\,(41)      & $-$0.21  & ~61 &  $-$61 & \, &  $-$0.19 &  ~79 &        & \, &           &     &	\\
 4369.41  &  Fe\,{\sc ii}\,(28)     & $-$0.15  & ~61 &        & \, &  $-$0.09 &  ~48 &        & \, &           &     &	\\
 4374.61  &  Ti\,{\sc ii}\,(93) +   & $-$0.26  & ~69 & $-$103 & \, &  $-$0.08 &  ~73 &        & \, &           &     &	\\
 4375.69  &  Sc\,{\sc ii}\,(14)\,(bl) &          &     &        & \, &          &      &        & \, &           &     &	\\
 4384.32  &  Fe\,{\sc ii}\,(32)     & $-$0.23  & ~64 &  $-$95 & \, &          &      &        & \, &           &     &	\\
 4385.38  &  Fe\,{\sc ii}\,(27)     & $-$0.37  & ~60 &        & \, &  $-$0.16 &  ~79 &        & \, &           &     &	\\
 4395.08  &  Ti\,{\sc ii}\,(19)     & $-$0.41  & ~74 & $-$145 & \, &  $-$0.16 &  126 &        & \, &           &     &	\\
 4416.83  &  Fe\,{\sc ii}\,(27) +   & $-$0.37  & ~60 &        & \, &  $-$0.17 &  ~78 &        & \, &           &     &	\\
 4417.73  &  Ti\,{\sc ii}\,(40)\,(bl) &          &     &        & \, &          &      &        & \, &           &     &	\\
 4468.49  &  Ti\,{\sc ii}\,(31) +   & $-$0.28  & ~51 & $-$114 & \, &  $-$0.14 &  ~69 &        & \, &           &     &	\\
 4469.15  &  Ti\,{\sc ii}\,(18)\,(bl) &          &     &        & \, &          &      &        & \, &           &     &	\\
 4489.18  &  Fe\,{\sc ii}\,(37)     & $-$0.49  & ~80 & $-$114 & \, &  $-$0.18 &  115 &        & \, &           &     &	\\
 4491.41  &  Fe\,{\sc ii}\,(37)     & $-$0.52  & ~57 &        & \, &  $-$0.14 &  138 &        & \, &           &     &	\\
 4501.37  &  Ti\,{\sc ii}\,(31)     & $-$0.39  & ~59 & $-$118 & \, &  $-$0.08 &  ~70 &        & \, &           &     &	\\
 4508.28  &  Fe\,{\sc ii}\,(38)     & $-$0.49  & ~39 &  $-$80 & \, &  $-$0.27 &  ~95 &        & \, &           &     &	\\
 4515.34  &  Fe\,{\sc ii}\,(37)     & $-$0.66  & ~69 &  $-$80 & \, &  $-$0.18 &  ~80 &        & \, &           &     &	\\
 4520.22  &  Fe\,{\sc ii}\,(37)     & $-$0.39  & ~51 &  $-$93 & \, &  $-$0.29 &  100 &        & \, &           &     &	\\
 4522.63  &  Fe\,{\sc ii}\,(38)     & $-$0.43  & ~63 &        & \, &  $-$0.4  &  100 &        & \, &           &     &	\\
 4533.97  &  Ti\,{\sc ii}\,(50) +   & $-$0.34  & ~59 &  $-$92 & \, &          &      &        & \, &           &     &	\\
 4534.17  &  Fe\,{\sc ii}\,(37)\,(bl) &          &     &        & \, &          &      &        & \, &           &     &	\\
 4541.52  &  Fe\,{\sc ii}\,(38)     & $-$0.27  & ~59 &        & \, &          &      &        & \, &           &     &	\\
 4549.47  &  Fe\,{\sc ii}\,(38) +   & $-$0.45  & ~56 &        & \, &  $-$0.32 &  ~80 &        & \, &           &     &	\\
 4549.62  &  Ti\,{\sc ii}\,(82)\,(bl) &          &     &        & \, &          &      &        & \, &           &     &	\\
 4555.89  &  Fe\,{\sc ii}\,(37)     & $-$0.41  & ~61 &        & \, &  $-$0.21 &  ~76 &        & \, &           &     &	\\
 4558.66  &  Cr\,{\sc ii}\,(44)     & $-$0.39  & ~56 &        & \, &  $-$0.26 &  ~76 &        & \, &           &     &	\\
 4571.09  &  Mg\,{\sc i}\,(1)   +   & $-$0.26  & ~51 &  $-$81 & \, &  $-$0.21 &  118 &        & \, &           &     &	\\
 4571.97  &  Ti\,{\sc ii}\,(82)\,(bl) &          &     &        & \, &          &      &        & \, &           &     &	\\
 4576.33  &  Fe\,{\sc ii}\,(38)     & $-$0.47  & ~72 &        & \, &  $-$0.31 &  134 &        & \, &           &     &	\\
 4583.83  &  Fe\,{\sc ii}\,(38)     & $-$0.61  & ~77 & $-$122 & \, &  $-$0.51 &  ~84 &        & \, &           &     &	\\
 4588.32  &  Cr\,{\sc ii}\,(44)     & $-$0.26  & ~55 &        & \, &  $-$0.19 &  ~92 &        & \, &           &     &	\\
 4629.34  &  Fe\,{\sc ii}\,(37)     & $-$0.62  & ~66 &        & \, &  $-$0.29 &  112 &        & \, &           &     &	\\
 4634.11  &  Cr\,{\sc ii}\,(44) +   & $-$0.14  & ~44 &        & \, &  $-$0.08 &  ~95 &        & \, &           &     &	\\
 4734.12  &  Fe\,{\sc ii}\,(25)\,(bl) &          &     &        & \, &          &      &        & \, &           &     &	\\
 4708.66  &  Ti\,{\sc ii}\,(49)     & $-$0.08  & ~54 &        & \, &  $-$0.05 &  ~61 &        & \, &           &     &	\\
 4731.44  &  Fe\,{\sc ii}\,(43)     & $-$0.31  & ~59 &        & \, &  $-$0.11 &  109 &        & \, &           &     &	\\
 4805.11  &  Ti\,{\sc ii}\,(92)     & $-$0.32  & ~52 &  $-$87 & \, &  $-$0.04 &  ~40 &        & \, &           &     &	\\
 4812.35  &  Cr\,{\sc ii}\,(30)     & $-$0.06  & ~51 &        & \, &  $-$0.08 &  ~64 &        & \, &           &     &	\\
 4814.53  &  [Fe\,{\sc ii}]         & $-$0.06  & ~55 &        & \, &  $-$0.08 &  ~64 &        & \, &           &     &	\\
 4824.13  &  Cr\,{\sc ii}\,(30)     & $-$0.15  & ~55 &        & \, &  $-$0.04 &  ~61 &        & \, &           &     &	\\
 4861.33  &  H$\beta$               & $-$5.92  & 110 & $-$965 & \, &  $-$3.03 &  123 & $-$880 & \, &           &     &	\\
 4876.41  &  Cr\,{\sc ii}\,(30)     & $-$0.11  & ~55 &        & \, &  $-$0.05 &  ~47 &        & \, &           &     &	\\
 4923.92  &  Fe\,{\sc ii}\,(42)     & $-$1.04  & ~85 & $-$186 & \, &  $-$0.52 &  102 & $-$54  & \, &           &     &	\\
 4993.36  &  Fe\,{\sc ii}\,(36)     & $-$0.39  & ~42 &        & \, &  $-$0.18 &  ~78 &        & \, &           &     &	\\
 5018.43  &  Fe\,{\sc ii}\,(42)     & $-$1.73  & 123 & $-$230 & \, &  $-$0.95 &  120 & $-$48  & \, &           &     &	\\
 5129.14  &  Ti\,{\sc ii}\,(86)     & $-$0.04  & ~33 &        & \, &  $-$0.04 &  ~45 &        & \, &           &     &	\\
 5132.67  &  Fe\,{\sc ii}\,(35)     & $-$0.33  & ~52 &        & \, &  $-$0.08 &  ~67 &        & \, &           &     &	\\
 5158.78  &  [Fe\,{\sc ii}]         & $-$0.18  & ~82 &        & \, &  $-$0.13 &  ~50 &        & \, &           &     &	\\
 5169.03  &  Fe\,{\sc ii}\,(42)     & $-$1.91  & 119 &        & \, &  $-$0.83 &  ~77 &        & \, &           &     &	\\
\hline
\end{tabular}
%\noindent
%\flushleft
\end{table*}
\setcounter{table}{2}
\begin{table*}
\centering
\caption{(Continued)}
\small
\begin{tabular}{@{}clccccccccccccc}
\hline
$\lambda$ [\AA] & Ident. &  \multicolumn{3}{c}{WHT, 01/23/2000}   && \multicolumn{3}{c}{BTA, 11/22/1991} &&  \multicolumn{3}{c}{WHT, 12/26/1996}\\
\noalign{\smallskip}
\cline{3-5} \cline{7-9} \cline{11-13}
\noalign{\smallskip} 
           &  & EW & FWHM & $v_{\rm out}$ & \, & EW & FWHM & $v_{\rm out}$ & \, & EW & FWHM & $v_{\rm out}$\\
           &  & [\AA] & [km~s$^{-1}$] & [km~s$^{-1}$] & \, & [\AA] & [km~s$^{-1}$] & [km~s$^{-1}$] & \, & [\AA] & [km~s$^{-1}$] & [km~s$^{-1}$]\\
\hline
 5188.71  &  Ti\,{\sc ii}\,(70)     & $-$0.26  & ~44 &        & \, &  $-$0.11 &  ~78 &        & \, &           &     &	\\
 5197.57  &  Fe\,{\sc ii}\,(49)     & $-$0.69  & ~66 &        & \, &  $-$0.51 &  130 &        & \, &           &     &	\\
 5211.54  &  Ti\,{\sc ii}\,(103)    & $-$0.17  & ~73 &        & \, &  $-$0.11 &  ~59 &        & \, &           &     &	\\
 5226.53  &  Ti\,{\sc ii}\,(70)     & $-$0.47  & ~66 &        & \, &  $-$0.06 &  ~52 &        & \, &  $-$0.16  & ~83 &	\\
 5234.62  &  Fe\,{\sc ii}\,(49)     & $-$0.44  & ~80 &        & \, &  $-$0.23 &  ~74 &        & \, &  $-$0.21  & ~59 &	\\
 5237.37  &  Cr\,{\sc ii}\,(43)     & $-$0.19  & ~58 &        & \, &  $-$0.09 &  ~81 &        & \, &  $-$0.09  & ~63 &	\\
 5261.62  &  [Fe\,{\sc ii}]         & $-$0.09  & ~58 &        & \, &  $-$0.05 &  ~58 &        & \, &           &     &	\\
 5269.54  &  Fe\,{\sc i}\,(15)      & $-$0.45  & ~91 &        & \, &          &      &        & \, &  $-$0.14  & ~60 &	\\
 5275.99  &  Fe\,{\sc ii}\,(49)     & $-$0.56  & ~58 &        & \, &  $-$0.25 &  ~88 &        & \, &  $-$0.15  & ~65 &	\\
 5279.88  &  Cr\,{\sc ii}\,(43)     & $-$0.11  & ~65 &        & \, &  $-$0.04 &  ~58 &        & \, &           &     &	\\
 5284.09  &  Fe\,{\sc ii}\,(41)     & $-$0.81  & ~65 &        & \, &  $-$0.16 &  ~73 &        & \, &  $-$0.07  & ~40 &	\\
 5305.85  &  Cr\,{\sc ii}\,(24)     & $-$0.14  & ~80 &        & \, &  $-$0.08 &  ~40 &        & \, &           &     &	\\
 5316.61  &  Fe\,{\sc ii}\,(49) +   & $-$1.24  & ~72 &        & \, &  $-$0.64 &  ~94 &        & \, &  $-$0.52  & ~91 &	\\
 5316.78  &  Fe\,{\sc ii}\,(48)\,(bl) &          &     &        & \, &          &      &        & \, &           &     &	\\
 5325.56  &  Fe\,{\sc ii}\,(49)     & $-$0.26  & ~72 &        & \, &  $-$0.11 &  ~58 &        & \, &   ~~0.04  & ~36 &	\\
 5328.04  &  Fe\,{\sc i}\,(15)      & $-$0.33  & ~67 &        & \, &          &      &        & \, &   ~~0.08  & ~45 &	\\
 5333.65  &  [Fe\,{\sc ii}]         & $-$0.07  & ~43 &        & \, &          &      &        & \, &  $-$0.08  & ~60 &	\\
 5336.81  &  Ti\,{\sc ii}\,(69)     & $-$0.35  & ~69 &        & \, &          &      &        & \, &   ~~0.11  & ~36 &	\\
 5362.86  &  Fe\,{\sc ii}\,(48)     & $-$0.71  & ~33 &        & \, &  $-$0.37 &  115 &        & \, &  $-$0.13  & ~43 &	\\
 5371.51  &  Fe\,{\sc i}\,(15)      & $-$0.17  & ~57 &        & \, &          &      &        & \, &  $-$0.07  & ~50 &	\\
 5397.13  &  Fe\,{\sc i}\,(15)      & $-$0.21  & ~59 &        & \, &          &      &        & \, &  $-$0.11  & ~47 &  \\
 5405.78  &  Fe\,{\sc i}\,(15)      & $-$0.15  & ~56 &        & \, &          &      &        & \, &  $-$0.06  & ~38 &	\\
 5429.67  &  Fe\,{\sc i}\,(15)      & $-$0.18  & ~47 &        & \, &          &      &        & \, &  $-$0.06  & ~46 &	\\
 5432.98  &  Fe\,{\sc ii}\,(55)     & $-$0.17  & ~66 &        & \, &          &      &        & \, &  $-$0.06  & ~53 &	\\
 5434.53  &  Fe\,{\sc i}\,(15)      & $-$0.08  & ~60 &        & \, &          &      &        & \, &  $-$0.02  & ~18 &	\\
 5446.92  &  Fe\,{\sc i}\,(15)      & $-$0.11  & ~47 &        & \, &          &      &        & \, &  $-$0.05  & ~28 &	\\
 5455.62  &  Fe\,{\sc i}\,(15)      & $-$0.14  & ~51 &        & \, &          &      &        & \, &  $-$0.04  & ~42 &	\\
 5497.53  &  Fe\,{\sc i}\,(15)      & $-$0.11  & ~65 &        & \, &          &      &        & \, &  $-$0.07  & ~46 &	\\
 5501.48  &  Fe\,{\sc i}\,(15)      & $-$0.06  & ~46 &        & \, &          &      &        & \, &  $-$0.03  & ~35 &	\\
 5506.79  &  Fe\,{\sc i}\,(15)      & $-$0.06  & ~69 &        & \, &          &      &        & \, &  $-$0.04  & ~46 &	\\
 5534.86  &  Fe\,{\sc ii}\,(55)     & $-$0.53  & ~64 &        & \, &          &      &        & \, &  $-$0.11  & ~48 &	\\
 5577.34  &  [O\,{\sc i}]           & $-$0.05  & ~73 &        & \, &          &      &        & \, &  $-$0.15  & ~37 &  \\
 5591.37  &  Fe\,{\sc ii}\,(55)     & $-$0.74  & ~68 &        & \, &          &      &        & \, &  $-$0.09  & ~36 &	\\
 5754.64  &  [N\,{\sc ii}]          & $-$0.05  & ~30 &        & \, &          &      &        & \, &  $-$0.02  & ~40 &  \\
 6562.81  &  H$\alpha$              & $-$44.1  & 191 & $-$920 & \, &  $-$16.8 &  200 & $-$850: & \, &  $-$18.6$^{\dagger}$  & ~96 & $-$1000\\
\hline
\end{tabular}
\noindent
\flushleft
$^{\dagger}$ Observed with ESO CAT, 12/16/1996.
\end{table*}
\begin{table*}
\centering
\caption{Lines detected in the red spectra of Z~CMa (bl = blend, s = single, tr = triplet).  Lines included in square brackets are forbidden transitions.}
\tabcolsep0.09cm
\small
\begin{tabular}{@{}clccccccccccccccccc}
\hline
$\lambda$ [\AA] & Ident. &  \multicolumn{3}{c}{Tautenburg, 01/23/2000}   && \multicolumn{3}{c}{WHT, 02/18/2000} &&  \multicolumn{3}{c}{NTT, 12/21/1991}  && \multicolumn{3}{c}{WHT, 12/26/1996}\\
\noalign{\smallskip}
\cline{3-5} \cline{7-9} \cline{11-13} \cline{15-17}
\noalign{\smallskip} 
           &  & EW & FWHM & $v_{\rm out}$ & \, & EW & FWHM & $v_{\rm out}$ & \, & EW & FWHM & $v_{\rm out}$ & \, & EW & FWHM & $v_{\rm out}$\\
           &  & [\AA] & [km~s$^{-1}$] & [km~s$^{-1}$] & \, & [\AA] & [km~s$^{-1}$] & [km~s$^{-1}$] & \, & [\AA] & [km~s$^{-1}$] & [km~s$^{-1}$] & \, & [\AA] & [km~s$^{-1}$] & [km~s$^{-1}$]\\
\hline
%  5642.11  &  Sc\,{\sc ii}\,(29)    &          &          &        & \, &          &         &       & \, &         &         &       & \, &           &       &         \\  
  5657.87  &  Sc\,{\sc ii}\,(29)    & $-$0.28  &    ~38   &        & \, &          &         &       & \, &         &         &       & \, &   ~~0.03  &  ~43  &         \\ 
%  5696.65  &  S\,{\sc i}\,(11)      & $-$0.05  &    ~12   &        & \, &          &         &       & \, &         &         &       & \, &           &       &         \\
  5875.62  &  He\,{\sc i}\,(tr)     &  ~~2.23  &    198   & $-$420 & \, &  ~~0.84  &    198  &$-$377 & \, &         &         &       & \, &           &       &         \\ 
  5889.95  &  Na\,{\sc i}\,(D$_1$) +&  ~~4.92  &    128   & $-$665 & \, &  ~~4.01  &    189  &$-$650 & \, &  ~~2.03 &    123  &$-$517 & \, &   ~~3.41  &   130 & $-$525  \\
  5895.92  &  Na\,{\sc i}\,(D$_2$)\,(bl) &       &          &        & \, &          &         &       & \, &         &         &       & \, &           &       &         \\ 
  5991.38  &  Fe\,{\sc ii}\,(46)    & $-$0.76  &    ~77   &        & \, &  $-$0.67 &    ~74 &       & \, &  $-$0.25 &    ~64  &       & \, &   $-$0.09 &   ~48 &         \\
  6084.11  &  Fe\,{\sc ii}\,(46)    & $-$0.29  &    109   &        & \, &  $-$0.37 &    100 &       & \, &  $-$0.07 &    ~63  &       & \, &           &       &         \\
  6113.33  &  Fe\,{\sc ii}\,(46)    & $-$0.06  &    ~55   &        & \, &          &        &       & \, &  $-$0.04 &    ~67  &       & \, &  	       &       &         \\
  6116.61  &  Fe\,{\sc ii}\,(46)    & $-$0.07  &    ~55   &        & \, &          &        &       & \, &          &         &       & \, & 	       &       &         \\
  6129.73  &  Fe\,{\sc ii}\,(46)    & $-$0.08  &    ~66   &        & \, &          &        &       & \, &          &         &       & \, &   $-$0.02 &   ~33 &         \\
  6136.33  &  Fe\,{\sc ii}\,(169)   & $-$0.13  &    ~42   &        & \, &          &        &       & \, &          &         &       & \, &           &       &         \\
  6147.74  &  Fe\,{\sc ii}\,(74) +  & $-$1.23  &          &        & \, &  $-$0.83 &        &       & \, &  $-$0.28 &         &       & \, &   $-$0.21 &       &         \\
  6149.25  &  Fe\,{\sc ii}\,(74)\,(bl)&          &          &        & \, &          &        &       & \, &          &         &       & \, &           &       &         \\
  6156.41  &  O\,{\sc i}\,(10)      &  ~~0.29  &          &        & \, &   ~~0.29 &   280  &       & \, &          &         &       & \, &           &       &         \\ 
  6191.75  &  Fe\,{\sc i}\,(169)    & $-$0.31  &    ~68   &        & \, &  $-$0.33 &    ~61 &       & \, &  $-$0.09 &    ~50  &       & \, &   $-$0.08 &   ~35 &         \\
  6238.39  &  Fe\,{\sc ii}\,(74)    & $-$0.99  &    ~94   & $-$360 & \, &  $-$0.65 &    ~86 &       & \, &  $-$0.15 &    ~73  &       & \, &   $-$0.02 &   ~31 &         \\
  6247.56  &  Fe\,{\sc ii}\,(74)    & $-$1.02  &    ~89   & $-$340 & \, &  $-$0.84 &    ~78 &       & \, &  $-$0.34 &    ~77  &       & \, & 	       &       &         \\
  6252.56  &  Fe\,{\sc i}\,(169)    & $-$0.31  &    ~68   &        & \, &  $-$0.15 &    ~53 &       & \, &          &         &       & \, &   $-$0.08 &   ~70 &         \\
  6300.31  &  [O\,{\sc i}]\,(1)     & $-$0.44  &          &        & \, &~$-$4.46$^{\dagger}$ &  &  & \, &  $-$0.36 &         &       & \, &  $-$2.63  &       &         \\ 
  6318.13  &  Fe\,{\sc i}\,(168)    & $-$0.27  &    ~40   &        & \, &  $-$0.13 &    ~54 &       & \, &  $-$0.06 &    ~52  &       & \, &           &       &         \\
  6331.95  &  Fe\,{\sc ii}\,(199)   & $-$0.08  &    ~88   &        & \, &  $-$0.06 &    ~60 &       & \, &  $-$0.08 &    ~64  &       & \, &  	       &       &         \\
  6347.09  &  Si\,{\sc ii}\,(2)     & $-$0.11  &          &        & \, &          &        &       & \, &          &         &       & \, &           &       &         \\  
  6363.79  &  [O\,{\sc i}]\,(1)     & $-$0.07  &          &        & \, &  $-$0.11 &    ~56 &       & \, &  $-$0.07 &    ~64  &       & \, &   $-$0.12 &   ~66 &         \\ 
  6369.46  &  Fe\,{\sc ii}\,(40)    & $-$0.18  &    ~61   &        & \, &  $-$0.15 &    ~49 &       & \, &  $-$0.15 &    ~52  &       & \, &  	       &       &         \\
  6371.36  &  Si\,{\sc ii}\,(2)     & $-$0.15  &    ~42   &        & \, &          &        &       & \, &          &         &       & \, &           &       &         \\  
  6393.61  &  Fe\,{\sc i}\,(168)    & $-$0.21  &    ~34   &        & \, &  $-$0.18 &    ~56 &       & \, &  $-$0.06 &    ~52  &       & \, &   $-$0.08 &   ~16 &         \\
  6407.29  &  Fe\,{\sc ii}\,(74)    & $-$0.13  &    109   &        & \, &  $-$0.12 &    104 &       & \, &  $-$0.07 &    ~60  &       & \, & 	       &       &         \\
  6416.91  &  Fe\,{\sc ii}\,(74)    & $-$0.69  &    ~85   &        & \, &  $-$0.37 &    ~69 &       & \, &  $-$0.2  &    ~61  &       & \, &   $-$0.09 &       &         \\
  6432.65  &  Fe\,{\sc ii}\,(40)    & $-$1.32  &    ~76   &        & \, &  $-$1.26 &    ~75 &       & \, &  $-$0.27 &    ~55  &       & \, &   $-$0.17 &   ~47 &         \\
  6456.38  &  Fe\,{\sc ii}\,(74)    & $-$1.36  &    ~85   &        & \, &          &        &       & \, &  $-$0.35 &    ~75  &       & \, &  	       &       &         \\
  6482.15  &  Fe\,{\sc ii}\,(199)   & $-$0.17  &    ~76   &        & \, &  $-$0.26 &    ~51 &       & \, &          &         &       & \, & 	       &       &         \\
  6516.81  &  Fe\,{\sc ii}\,(40)    & $-$1.78  &    ~84   & $-$330 & \, &  $-$1.28 &    ~60 &       & \, &  $-$0.58 &    ~74  &       & \, &   $-$0.29 &   ~43 &         \\
  6548.04  &  [N\,{\sc ii}]         & $-$0.03  &    ~17   &        & \, &          &        &       & \, &          &         &       & \, &           &       &         \\
  6562.81  &  H$\alpha$             & $-$42.4  &    200   &$-$1100 & \, & $-$44.1  &    191  &$-$920 & \, & $-$17.9 &    120  & $-$1040 & \, &  $-$23.4  &   ~96 &$-$1000  \\
  6583.46  &  [N\,{\sc ii}]         & $-$0.02  &    ~18   &        & \, & $-$0.02  &    ~22  &       & \, & $-$0.02 &    ~23  &       & \, &  $-$0.03  &   ~17 &         \\ 
  6678.15  &  He\,{\sc i}\,(s)      &  ~~0.67  &    300   &        & \, &          &         &       & \, &  ~~0.06 &    ~53  &       & \, &           &       &         \\
           &  He\,{\sc i}\,(s)      & $-$0.03  &          &        & \, &          &         &       & \, &         &         &       & \, &           &       &         \\  
  6707.81  &  Li\,{\sc i}\,(1)      &  ~~0.06  &    ~28   &        & \, &          &         &       & \, &  ~~0.05 &    ~28  &       & \, &   ~~0.09  &       &         \\ 
  6717.24  &  [S\,{\sc ii}]         & $-$0.13  &    ~20   &        & \, & $-$0.16  &    ~60  &       & \, &         &         &       & \, &  $-$0.03  &       &         \\ 
  6726.28  &  O\,{\sc i}\,(2)       & $-$0.02  &    ~75   &~$-$200 & \, &          &         &       & \, &         &         &       & \, &           &       &         \\
  6731.31  &  [S\,{\sc ii}]         & $-$0.07  &    ~20   &        & \, & $-$0.12  &         &       & \, & $-$0.03 &         &       & \, &  $-$0.05  &       &         \\ 
  6757.21  &  S\,{\sc i}\,(8)       & $-$0.01  &    ~33   &        & \, &          &         &       & \, &         &         &       & \, &           &       &         \\ 
  7005.89  &  Si\,{\sc i}\,(60)     & $-$0.05  &    ~33   & $-$110 & \, & $-$0.04  &    ~58  &       & \, &         &         &       & \, &           &       &         \\ 
  7017.67  &  Si\,{\sc i}\,(51)     & $-$0.04  &    ~44   & $-$110 & \, &          &         &       & \, &         &         &       & \, &           &       &         \\ 
  7065.19  &  He\,{\sc i}\,(tr)     & $-$0.44  &    280   & $-$430 & \, & $-$0.07  &    590  &       & \, & $<$0.04 &         &       & \, &           &       &         \\  
  7155.16  &  [Fe\,{\sc ii}]        &          &          &        & \, & $-$0.10  &    ~50  &       & \, & $-$0.10 &    ~85  &       & \, &           &       &         \\
  7244.85  &  Si\,{\sc i}\,(51)     & $-$0.04  &    ~44   & $-$110 & \, &          &         &       & \, & $-$0.10 &         &       & \, &           &       &         \\ 
  7281.35  &  He\,{\sc i}\,(s)      & $-$0.65  &    ~30   &        & \, &          &         &       & \, & $-$0.07 &         &       & \, &           &       &         \\ 
           &  He\,{\sc i}\,(s)      &  ~~0.14  &          &        & \, &          &         &       & \, &         &         &       & \, &           &       &         \\
  7291.46  &  [Ca\,{\sc ii}]\,(1)   & $-$0.51  &    ~56   &  $-$77 & \, &          &         &       & \, & $-$0.21 &    ~63  &       & \, &  $-$0.21  &   ~21 &         \\ 
  7310.21  &  Fe\,{\sc ii}\,(73)    & $-$0.35  &  $<$95   &        & \, &          &        &       & \, &  $-$0.11 &    ~69  &       & \, &  	       &       &         \\
  7320.69  &  Fe\,{\sc ii}\,(73)    & $-$0.27  &    ~55   &        & \, &          &        &       & \, &  $-$0.11 &    ~66  &       & \, &  	       &       &         \\
  7323.88  &  [Ca\,{\sc ii}]\,(1)   & $-$0.45  &    ~58   &  $-$77 & \, &          &         &       & \, & $-$0.11 &    ~38  &       & \, &  $-$0.14  &   ~30 &         \\ 
  7377.91  &  [Ni\,{\sc ii}]\,(2)   & $-$0.04  &    ~24   &  $-$60 & \, & $-$0.01  &    ~27  &       & \, &         &         &       & \, &           &       &         \\ 
  7411.87  &  [Ni\,{\sc ii}]\,(2)   & $-$0.08  &    ~64   &        & \, &          &         &       & \, & $-$0.03 &    ~31  &       & \, &  $-$0.05  &   ~55 &         \\
  7415.96  &  S\,{\sc i}\,(22)      & $-$0.19  &    100   & $-$100 & \, &  $-$0.15 &    ~87  &       & \, & $-$0.04 &    ~51  &       & \, &           &       &         \\
  7449.34  &  Fe\,{\sc ii}\,(73)    & $-$0.32  &    ~86   &        & \, &  $-$0.28 &    ~78 &       & \, &  $-$0.15 &    ~90  &       & \, &  	       &       &         \\
  7462.34  &  Fe\,{\sc ii}\,(73)    & $-$0.71  &    ~68   &        & \, &  $-$0.57 &    ~72 &       & \, &  $-$0.14 &    ~48  &       & \, &  	       &       &         \\
  7515.84  &  Fe\,{\sc ii}\,(73)    & $-$0.25  &    ~60   &        & \, &  $-$0.22 &    ~75 &       & \, &  $-$0.08 &    ~47  &       & \, &   $-$0.03 &   ~71 &         \\
  7698.96  &  K\,{\sc i}\,(1)       &  ~~0.08  &    ~12   &        & \, &  ~~0.32  &    ~15  &       & \, &  ~~0.20 &         &       & \, &    ~~0.17 &       &         \\ 
           &  K\,{\sc i}\,(1)       & $-$0.08  &          &        & \, &  $-$0.09 &         &       & \, & $-$0.05 &         &       & \, &   $-$0.08 &       &         \\
  7711.73  &  Fe\,{\sc ii}\,(73)    & $-$1.11  &    ~82   &        & \, &  $-$0.9  &    ~76 &       & \, &  $-$0.25 &    ~66  &       & \, &   $-$0.06 &   ~50 &         \\
\hline
\end{tabular}
%\noindent
%\flushleft
\end{table*}
\setcounter{table}{3}
\begin{table*}
\centering
\caption{(Continued)}
\tabcolsep0.09cm
\small
\begin{tabular}{@{}clccccccccccccccccc}
\hline
$\lambda$ [\AA] & Ident. &  \multicolumn{3}{c}{Tautenburg, 01/23/2000}   && \multicolumn{3}{c}{WHT, 02/18/2000} &&  \multicolumn{3}{c}{NTT, 12/21/1991}  && \multicolumn{3}{c}{WHT, 12/26/1996}\\
\noalign{\smallskip}
\cline{3-5} \cline{7-9} \cline{11-13} \cline{15-17}
\noalign{\smallskip} 
           &  & EW & FWHM & $v_{\rm out}$ & \, & EW & FWHM & $v_{\rm out}$ & \, & EW & FWHM & $v_{\rm out}$ & \, & EW & FWHM & $v_{\rm out}$\\
           &  & [\AA] & [km~s$^{-1}$] & [km~s$^{-1}$] & \, & [\AA] & [km~s$^{-1}$] & [km~s$^{-1}$] & \, & [\AA] & [km~s$^{-1}$] & [km~s$^{-1}$] & \, & [\AA] & [km~s$^{-1}$] & [km~s$^{-1}$]\\
\hline
  7774.17  &  O\,{\sc i}\,(1)(tr)   &  ~~2.79  &   ~340   &   ~450 & \, &  ~~3.12  &    300  &~$-$250& \, &  ~~0.11 &         &       & \, &           &       &         \\ 
           &  O\,{\sc i}\,(1)       & $-$0.08  &          &        & \, & $-$0.19  &         &       & \, &         &         &       & \, &           &       &         \\
  7877.51  &  Mg\,{\sc ii}\,(8)     & $-$0.18  &    ~80   & $-$304 & \, &          &         &       & \, &         &         &       & \, &           &       &         \\ 
  7896.38  &  Mg\,{\sc ii}\,(8)     & $-$0.23  &          & $-$304 & \, &          &         &       & \, & $-$0.03 &         &       & \, &           &       &         \\ 
  7918.38  &  Si\,{\sc i}\,(57)     & $-$0.08  &    ~66   &             &          &         &       & \, &         &         &       & \, &           &       &         \\
  7932.35  &  Si\,{\sc i}\,(57)     & $-$0.08  &    ~66   &             &          &         &       & \, &         &         &       & \, &           &       &         \\
  7944.11  &  Si\,{\sc i}\,(57)     & $-$0.07  &    ~55   &        & \, &          &         &       & \, & $-$0.06 &         &       & \, &           &       &         \\ 
  8000.12  & [Cr\,{\sc ii}]\,(1)    & $-$0.21  &    ~60   &        & \, &          &         &       & \, &         &         &       & \, &           &       &         \\ 
  8012.27  &  S\,{\sc i}\,(10)      & $-$0.04  &    ~44   &        & \, &          &         &       & \, &         &         &       & \, &           &       &         \\ 
  8047.81  &  Mg\,{\sc i}\,(0)      & $-$0.17  &    ~66   & $-$250 & \, &          &         &       & \, &         &         &       & \, &  $-$0.05  &   ~32 &	 \\
  8201.71  &  Ca\,{\sc ii}\,(13)    & $-$0.15  &    ~60   &        & \, & $-$0.18  &    ~87  &       & \, & $-$0.05 &         &       & \, &  $-$0.05: &       &         \\
  8213.99  &  Mg\,{\sc ii}\,(7)     & $-$0.03  &    ~60   &        & \, & $-$0.07  &    ~78  &       & \, &         &         &       & \, &           &       &         \\ 
  8234.64  &  Mg\,{\sc ii}\,(7)     & $-$0.14  &    ~60   &        & \, & $-$0.22  &    ~82  &       & \, & $-$0.14 &    ~42  &       & \, &           &       &         \\ 
  8248.81  &  Ca\,{\sc ii}\,(13)    & $-$0.27  &    ~64   & $-$550 & \, & $-$0.27  &    ~83  &$-$250 & \, &         &         &       & \, &           &       &         \\ 
  8327.11  &  Fe\,{\sc i}\,(60)     & $-$0.26  &    ~60   &        & \, & $-$0.27  &    ~76  &       & \, & $-$0.12 &    ~51  &       & \, &           &       &         \\
  8345.45  &  P23 +                 & $-$0.53  &    120   &        & \, &          &         &       & \, & $-$0.11 &    ~39  &       & \, &           &       &         \\ 
  8346.13  &  Mg\,{\sc i}\,(40)\,(bl) &          &          &        & \, &          &         &       & \, &         &         &       & \, &           &       &         \\  
  8359.00  &  P22 +                 & $-$0.47  &    128   &        & \, &          &         &       & \, & $-$0.19 &    110  &       & \, &           &       &         \\
  8361.69  &  He\,{\sc i}\,(bl)     &          &          &        & \, &          &         &       & \, &         &         &       & \, &           &       &         \\  
  8374.48  &  P21                   & $-$0.44  &    ~80   &        & \, &          &         &       & \, &         &         &       & \, &           &       &         \\ 
  8387.78  &  Fe\,{\sc i}\,(60)     & $-$0.64  &    ~80   &        & \, &          &         &       & \, & $-$0.19 &    ~58  &       & \, &  $-$0.19  &   ~60 &         \\
  8392.40  &  P20                   & $-$0.63  &    100   &        & \, &          &         &       & \, &         &         &       & \, &           &       &         \\ 
  8413.32  &  P19                   & $-$0.54  &    110   & $-$440 & \, &          &         &       & \, &         &         &       & \, &           &       &         \\ 
%  8437.98  &  P18                   &          &          &        & \, &          &         &       & \, &         &         &       & \, &           &       &         \\  
  8446.51  &  O\,{\sc i}\,(4)       & $-$1.37  &    107   &        & \, &  ~~1.82  &    165  &$-$270 & \, & $-$1.31 &    120  &       & \, &           &       &         \\ 
  8467.25  &  P17(bl)               & $-$0.58  &    ~96   & $-$430 & \, & $-$0.68  &    111  &$-$255 & \, & $-$0.17 &    ~75  &       & \, &           &       &         \\ 
  8468.42  &  Fe\,{\sc i}\,(60)     &          &          &        & \, &          &         &       & \, &         &         &       & \, &           &       &         \\ 
  8498.02  &  Ca\,{\sc ii}\,(2)     & $-$12.6  &    107   &        & \, &$-$10.86  &    ~83  &$-$430 & \, & $-$5.51 &    ~93  &$-$250 & \, &  $-$4.76  &   101 &  $-$282 \\ 
  8502.41  &  P16(bl)               &          &          &        & \, &          &         &       & \, &         &         &       & \, &           &       &         \\  
  8514.10  &  Fe\,{\sc i}\,(60)     & $-$0.48  &    ~62   &        & \, & $-$0.38  &    ~66  &       & \, & $-$0.07 &    ~60  &       & \, &  $-$0.05  &   ~44 &         \\ 
  8542.10  &  Ca\,{\sc ii}\,(2) +   & $-$14.9: &    110   & $-$500 & \, & $-$11.3  &    ~81  &$-$460 & \, & $-$5.82 &    109  &$-$320 & \, &  $-$4.95  &   ~79 &  $-$307 \\ 
  8545.38  &  P15(bl)               &          &          &        & \, &          &         &       & \, &         &         &       & \, &           &       &         \\  
  8598.39  &  P14                   & $-$0.61  &    ~89   & $-$420 & \, & $-$0.37  &    ~83  &       & \, & $-$0.19 &    ~31  &       & \, &           &       &         \\ 
  8629.24  &  N\,{\sc i}\,(2)       & $-$0.17  &          &        & \, &          &         &       & \, &         &         &       & \, &           &       &         \\  
  8662.14  &  Ca\,{\sc ii}\,(2) +   & $-$11.8  &    132   & $-$440 & \, & $-$14.9  &    ~73  &$-$500 & \, & $-$5.26 &    106  &$-$277 & \, &  $-$3.59  &   109 &  $-$430 \\ 
  8665.02  &  P13(bl)               &          &          &        & \, &          &         &       & \, &         &         &       & \, &           &       &         \\  
  8683.41  &  N\,{\sc i}\,(1)       & $-$0.06  &    ~55   &        & \, &          &         &       & \, & $-$0.03 &    ~10  &       & \, &           &       &         \\ 
  8688.64  &  Fe\,{\sc i}\,(60)     & $-$0.65  &    ~76   &        & \, &          &         &       & \, & $-$0.12 &    ~50  &       & \, &  $-$0.08  &   ~50 &         \\
  8728.89  &  N\,{\sc i}\,(1)       & $-$0.16  &    ~56   &        & \, &          &         &       & \, & $-$0.18 &    ~20  &       & \, &           &       &         \\ 
  8750.47  &  P12                   & $-$0.64  &    ~93   & $-$430 & \, & $-$0.29  &     87  &       & \, & $-$0.32 &    ~64  &       & \, &           &       &         \\ 
  8806.74  &  Mg\,{\sc i}\,(7)      & $-$0.84  &    111   & $-$325 & \, & $-$0.83  &    101  &       & \, & $-$0.06 &    ~35  &       & \, &  $-$0.04  &   ~20 &         \\ 
  8824.11  &  Fe\,{\sc i}\,(60)     & $-$0.51  &    110   &        & \, & $-$0.48  &    ~66  &       & \, & $-$0.08 &    ~66  &       & \, &  $-$0.05  &       &         \\ 
  8862.78  &  P11                   & $-$0.64  &    ~85   & $-$433 & \, & $-$0.51  &    ~85  &       & \, & $-$0.28 &    ~52  &       & \, &           &       &         \\ 
  8912.07  &  Ca\,{\sc ii}          & $-$0.58  &    106   & $-$415 & \, &          &         &       & \, & $-$0.05 &    ~40  &       & \, &           &       &         \\
  8927.36  &  Ca\,{\sc ii}          & $-$0.82  &    106   & $-$420 & \, &          &         &       & \, & $-$0.06 &    ~51  &       & \, &           &       &         \\ 
  9014.91  &  P10                   &          &          &        & \, &          &         &       & \, &  ~~0.32 &    ~54  &       & \, &           &       &         \\ 
  9063.27  &  He\,{\sc i}\,(77)     & $-$0.68  &    150   & $-$435 & \, &          &         &       & \, &         &         &       & \, &           &       &         \\ 
  9229.02  &  P9                    & $-$0.61  &    ~95   & $-$410 & \, &          &         &       & \, &  ~~0.49 &    ~66  &       & \, &           &       &         \\ 
  9545.97  &  P8                    & $-$0.82  &    ~90   & $-$440 & \, &          &         &       & \, &         &         &       & \, &           &       &         \\ 
  9854.74  &  Ca\,{\sc ii}          & $-$0.43  &    106   &        & \, &          &         &       & \, &         &         &       & \, &           &       &         \\ 
%  9956.31  &  Fe\,{\sc ii}          &          &          &        & \, & $-$0.06  &         &       & \, &         &         &       & \, &           &       &         \\
  9997.57  &  Fe\,{\sc ii}          & $-$2.02  &    110   & $-$360 & \, & $-$1.71  &    110  &$-$380 & \, &  ~~0.00 &         &       & \, &   ~~0.00  &       &         \\
 10049.38  &  P7                    & $-$1.1   &    ~88   & $-$420 & \, & $-$0.85  &    ~84  &$-$342 & \, &  ~~0.00 &         &       & \, &   ~~0.00  &       &         \\
 10370.51  &  [S\,{\sc ii}]         &          &          &        & \, & $-$0.08  &         &       & \, &         &         &       & \, &           &       &         \\
 10397.74  &  [N\,{\sc i}]          &          &          &        & \, & $-$0.16  &         &       & \, &         &         &       & \, &           &       &         \\
% 10404.17  &  [N\,{\sc i}]          &          &          &        & \, & $-$0.08  &         &       & \, &         &         &       & \, &           &       &         \\
 10459.79  &  [Ni\,{\sc ii}]        &          &          &        & \, & $-$0.31  &         &       & \, &         &         &       & \, &           &       &         \\
 10501.50  &  Fe\,{\sc ii}          &          &          &        & \, & $-$0.09  &         &       & \, &         &         &       & \, &           &       &         \\
% 10514.50  &  P\,{\sc i}/N\,{\sc i} &          &          &        & \, & $-$0.13  &         &       & \, &         &         &       & \, &           &       &         \\
 10525.14  &  Fe\,{\sc ii}          &          &          &        & \, & $-$0.16  &         &       & \, &         &         &       & \, &           &       &         \\
 10931.83  &  P\,{\sc i}            &          &          &        & \, & $-$0.23  &         &       & \, &         &         &       & \, &           &       &         \\
 10937.83  &  P6 (P$\gamma$)        &          &          &        & \, & $-$0.10  &         &       & \, &         &         &       & \, &           &       &         \\
\hline
\end{tabular}
\noindent
\flushleft
$^{\dagger}$ Observed with ESO 3.6 m, 03/31/2002.
\end{table*}

\subsection{The spectra of  Nov./Dec. 1991 ($V$ = 9.6\mag)}
\subsubsection{The emission lines}
\begin{figure}
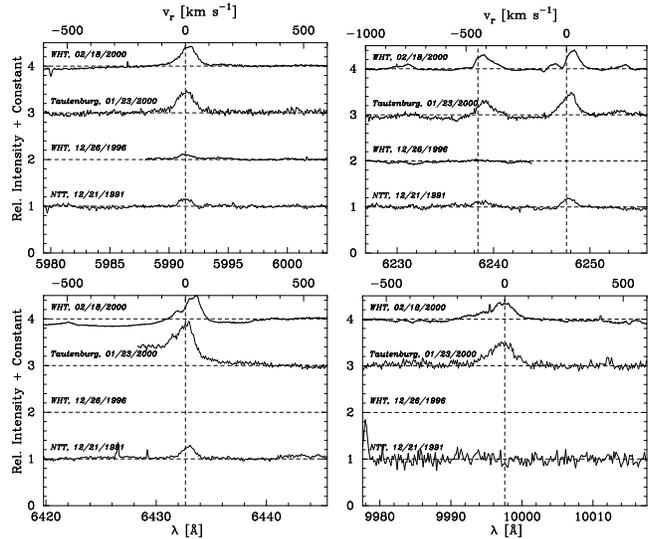

\vspace*{0.15cm}
\centerline{\psfig{figure=zcma_fig7a.ps,width=4.2cm,angle=270}
            \hspace*{0.05cm}
            \psfig{figure=zcma_fig7b.ps,width=4.0cm,angle=270}}
\vspace*{0.1cm}
\centerline{\psfig{figure=zcma_fig7c.ps,width=4.2cm,angle=270}
            \hspace*{0.05cm}
            \psfig{figure=zcma_fig7d.ps,width=4.0cm,angle=270}}
\caption[]{Fe\,{\sc ii} line profile variations in Z~CMa between 
1991 and 2000.  (a) Fe\,{\sc ii}\,(46) 5991.38\,\AA\ line profile 
variations, (b) the same for the Fe\,{\sc ii}\,(74) 6238.38, 6247.56\,\AA\ 
lines, (c) Fe\,{\sc ii}\,(40) 6432.65\,\AA\ line, (d) Fe\,{\sc ii} 
9997.58 line.  Note the marked increase in Fe\,{\sc ii} line strength 
in the Jan. and Feb. 2000 spectra.}
\end{figure}
Although for Z~CMa indications for spectral changes related with
brightness variation have been reported previously in the literature 
(Covino et al. 1984; Finkenzeller \& Jankovics 1984; Hessman et al. 1991;
Hamann \& Persson 1992b; Welty et al. 1992; Teodorani et al. 1997; Garcia
et al. 1999), a systematic comparison of red and blue spectra of 
Z~CMa in its ``low'' and ``high''-state has not been possible so far.  
Using previously unpublished spectra from Z~CMa in its ``low state'', 
taken from the ESO data-archive, as well as our new data of Z~CMa after its 
Jan. 2000 outburst, we can now compare spectra of Z~CMa in its two states 
over the full spectral range between 4200 and 10000\,\AA. 

From this comparison we draw the following conclusions:
(1) The EWs of the Paschen and Ca\,{\sc ii}\,(2)  emission components in the
Jan. 2000 spectra are a factor
of 2.0--2.5 larger than the corresponding EWs in the Dec. 1991
spectra. We discuss the consequences
of this in Sect. 4.5. The half-widths of the Ca\,{\sc ii} lines are comparable
to those of Jan. 2000 but the outflow velocities are 140--180 km~s$^{-1}$
lower than in 2000. 
%No Ca\,{\sc ii}\,K profile is present within the range of the 1991 spectra.
(2) The EW of the O\,{\sc i} (8446\,\AA) emission line does not seem to increase
significantly between the observations in 1991 and 2000 but it is 
$\sim$ 16\% larger in Dec. 1987 (Hamann \& Persson 1992a) than in Jan. 2000. 
If this line is excited by Ly$\beta$ pumping, it indicates that it is formed 
at a location where the Ly$\beta$ fluxes
are not significantly different for Dec. 1991 and Jan. 2000.
(3) The weak N\,{\sc i} emission lines at 8728 and 8683\,\AA, seen in the Jan. 
2000 spectra, are also present in the 1991 NTT spectrum.
(4) All obtained spectra contain many emission lines of Fe\,{\sc ii}, Ti\,{\sc ii} 
and Cr\,{\sc ii}. The analysis of the Fe\,{\sc ii} lines
is given in Sect. 4.6, together with that of the outburst spectra of
1987 and 2000.  We also observe the emission lines of  Fe\,{\sc i}\,(60).
(5) In contrast to the other Fe\,{\sc ii} emission lines, the Fe\,{\sc ii} (9997\,\AA)
 line, which is extremely strong in the spectra of Jan./Feb. 2000, is not
 present in the NTT spectrum of 21 Dec. 1991. This line is the strongest
line in the near-IR multiplet $b^4G - z^4F^0$ from which four lines are
observed in the spectra of Jan./Feb.
2000 (Table 4) and in which the 9997\,\AA\ line is by far the strongest
transition. The lines may be pumped
directly from Ly$\alpha$ (fluorescence), by cascading from other 
Ly$\alpha$-pumped levels, or collisionally pumped
from the metastable $a^4G$ level. These three possible mechanisms
correspond with different excitations
in the UV (Rodr\'{\i}guez-Ardila et al 2002). Without high resolution UV
spectra of Z~CMa the method of formation cannot be selected.

\subsubsection{The absorption lines}
The strong and broad He\,{\sc i} (7065, 6678 and 5876\,\AA) and 
O\,{\sc i} (7773 and 6156\,\AA) absorption troughs
in the spectra of Jan./Feb. 2000 are much less prominent in the 
spectra of Dec. 1991.  For these lines the same variation was observed by 
Hessman et al. (1991)
between the `high state' and the `post-eruption state'. It suggests
that the contribution by the FUor dominates the local continuum in
Dec. 1991. This is probably also the reason, why the He\,{\sc i} (5876\,\AA)
absorption is not present in the Na\,{\sc i}\,D spectrum of Dec. 1991 and 
Dec. 1996 (Fig.~5).  We also find a corresponding reduction in the strength of the
He\,{\sc i} 7068\,\AA\ line, together with the appearance of narrow absorption
lines of Fe\,{\sc i} at 7068\,\AA\ and of
Fe\,{\sc i}, Ti\,{\sc i}, Mn\,{\sc i} at 7069--7070\,\AA\ and Fe\,{\sc i} at 7071\,\AA. 
The shape of the O\,{\sc i} (7773\,\AA) profile is shallow around 7772\,\AA\ and
shows a very small emission peak around 7776\,\AA. 
The blue wing of the 5889.95 Na\,{\sc i}\,D$_1$ line in the 2000, 1996 and 1991
spectra changes in shape. The first absorption component ($v$ $<$ 130 km~s$^{-1}$) 
increases in depth between
1991 and Feb. 2000, but decreases between 1996 and 1991. However, the
Jan. 2000 spectrum shows a second absorption wing with $v$ $<$ 380 km~s$^{-1}$, 
which suggests that both lines are formed in the variable outer parts of the 
envelope or disc atmosphere. 
The K\,{\sc i} (7699\,\AA) line profile is similar to that of the WHT
spectra of Feb. 2000.

\subsection{The spectra of Dec. 1996 ($V$ = 10.2\mag)}
\subsubsection{The emission lines}
   The red WHT spectrum of Dec. 1996, the ESO echelle profiles and
the H$\alpha$ profile of Bailey (1998) have been obtained during the deep
photometric minimum ($V$ $\sim$ 10.2\mag) between 1996 and 1999 (Fig. 1). 
Parts of  similar spectra, taken in Jan. 1996  and Jan. 1997 have been 
published by Teodorani et al. (1997) and Chochol et al. (1998) respectively. 
Our low resolution spectra of Dec. 1992 also correspond to this brightness.
Since the brightness is changing slowly during these years, we have no
direct indications for the relative
spectral contributions of the components of Z~CMa for this phase.
However, some hints can be
given by comparing these spectra with those of 1991, 2000 and
Feb. 1987 (see next section).
(1) The ESO profile of H$\alpha$ of Dec. 12 appears similar to the
 WHT profiles of Dec. 26. The EW
of the H$\alpha$ emission has increased slightly with respect to that of
1991. However, Chochol et
al. (1998) have found that the daily variations of the H$\alpha$ emission can
be considerable so that small changes
over long periods do not permit us to draw any conclusions. We also
note here that the Paschen emission
lines are not detected in the 1996 spectrum, so that the Ca\,{\sc ii}\,(2) lines
do not have to be corrected for the contribution by the Paschen lines.
(2) The ESO profiles of the Ca\,{\sc ii}\,(2) lines at 8498 and 8542\,\AA\ are very
similar to those of the WHT spectrum of Dec. 26. The EWs of the
strong emission lines of Ca\,{\sc ii}\,(2) and Fe\,{\sc ii} are lower by a factor of $\sim$
0.6 than the corresponding EWs  measured from the 1991  spectrum.
We discuss the consequences for the radii of the emission region in
sections 4.5 and 4.6. The profiles of the Ca\,{\sc ii} lines are similar to
those of Dec. 1991.
(3) The EW of the Fe\,{\sc i}\,(60) line at 8387\,\AA\ is reduced by a factor  of
0.85.
(4) The EWs of the forbidden emission line of O\,{\sc i} (6300\,\AA) have
gradually increased since 1987. For [Fe\,{\sc ii}] 7155\,\AA\ 
and the [Ca\,{\sc ii}] lines at 7291 and 7324\,\AA\ there also may be an 
increase with respect to the 1991 data. A probable explanation for this 
rise is given in Sect. 4.6.
(5) The EW of the Ca\,{\sc ii}\,K emission varies less than those of the
 Ca\,{\sc ii}\,(2) triplet. It even slightly
decreases between 1996 and the outburst stages of 1987 and 2000. This
can possibly be due to the
fact that the Ca\,{\sc ii}\,K line is formed in the outer envelope, in contrast
to the Ca\,{\sc ii}\,(2) disc lines.
\begin{figure*}
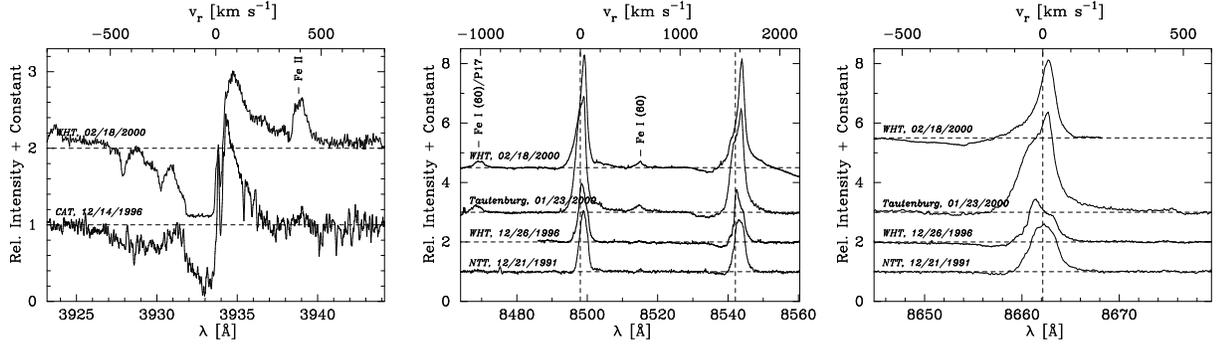

\vspace*{0.15cm}
\centerline{\psfig{figure=zcma_fig8a.ps,height=4.5cm,angle=270}
            \hspace*{0.3cm}
            \psfig{figure=zcma_fig8b.ps,height=4.5cm,angle=270}
            \hspace*{0.08cm}
            \psfig{figure=zcma_fig8c.ps,height=4.5cm,angle=270}}
\caption[]{Ca\,{\sc ii} line profile variations in Z~CMa between 
1991 and 2000.  From left to right: (a) the Ca\,{\sc ii}\,K line 
at 3933.66\,\AA\ (this spectral range includes Fe\,{\sc ii} 3938.29\,\AA, 
seen in emission in the Feb. 2000 spectrum), (b) the 8498.03 and 
8542.09\,\AA\ line of the red Ca\,{\sc ii}\,(2) triplet (this 
spectral range also includes the 8514.08\,\AA\ Fe\,{\sc i} line, 
and the 8467.8\,\AA\ P17/Fe\,{\sc i} blend), and 
(c) the 8662.14\,\AA\ line of the Ca\,{\sc ii}\,(2) triplet 
(blended with P13). Again 
note the marked increase in line strength in the Jan. and 
Feb. 2000 spectra.}
\end{figure*}

\subsubsection{The absorption lines}
(1) Although the He\,{\sc i} (7065\,\AA) line has not been observed in the high
state of Feb. 1987 and although we observe only the blue half of the
He\,{\sc i} (6678\,\AA) line in the 1996 spectrum, we can compare the remaining
profiles of these lines in the available spectra. We find that the
EWs of these lines are decreasing with
decreasing brightness, which indicates a decreasing contribution of
the primary B0e star. We can
extrapolate this increasing trend and find that in Dec. 1996 the
contribution of the primary is at most 10\%.
We would expect a corresponding increase of the EW for the 
Li\,{\sc i} (6707\,\AA) line (which is only contributed
by the FUor component), but the relatively large errors in line 
flux due to the weakness of this line in the four
spectra prohibit drawing any reliable conclusions regarding 
an increase in its EW. The decreasing trend is also found for the 
He\,{\sc i} (5876\,\AA) absorption line. In the 1996 spectrum
this line is too weak to be distinguishable in the neighbourhood of
the Na\,{\sc i}\,D absorption doublet. 
(2) The O\,{\sc i} (7773\,\AA) has not been observed in this spectrum (it lies
in the gap between order 36 and 37). 
(3) The Na\,{\sc i}\,D profiles of the ESO and WHT spectra of Dec. 1996 
are similar, but the WHT spectrum of
Dec. 26 has a higher resolution and shows that the blue components of
the lines are double with mean velocities of $-$53 and $-$128 km~s$^{-1}$. 
The Na\,{\sc i}\,D and K\,{\sc i} profiles do not differ much from those observed in 1991. 
The comparison of our 1996 Ca\,{\sc ii}\,K profile can be made only with those of the
outburst spectra of 1987 and 2000. Our 1991 profile has a higher
emission component than the other two profiles.  The blue absorption
wings of the three profiles are very similar, but in the 2000 spectrum
we observe two narrow absorption components at $-$305 and $-$460 km~s$^{-1}$,
which are difficult to find in the other two spectra, because of the
noise.

\subsection{The Ca\,{\sc ii}\,(2) emission lines}
\begin{table}
\centering
\caption{Ca\,{\sc ii}\,(2) line-ratios relative to the 8542\,\AA\ line in the 
spectra of Z~CMa, V645 Cyg, FU Ori and V1057 Cyg.}
\tabcolsep0.03cm
\small
\begin{tabular}{@{}lccccc}
\hline              
                    & \multicolumn{2}{c}{EW}  & \, &  \multicolumn{2}{c}{Peak Height} \\
\noalign{\smallskip}
\cline{2-3} \cline{5-6}
\noalign{\smallskip} 
                    & 8498/8542 & 8662/8542 && 8498/8542 & 8662/8542\\
\hline 
Z~CMa (Jan. 2000)   &  0.85  &  0.79  & \, &  1.44  &  0.73\\
Z~CMa (Feb. 2000)   &  0.96  &  --    & \, &  1.06  &  --  \\
Z~CMa (Dec. 1991)   &  0.95  &  0.90  & \, &  1.18  &  0.94\\
Z~CMa (Dec. 1996)   &  0.96  &  0.73  & \, &  1.10  &  0.96\\
Z~CMa (Nov. 1988)$^{\dagger}$   &  0.87  &  0.79  & \, &  1.00  &  0.86\\
V645 Cyg$^{\ddagger}$           &  0.86  &  0.95  & \, &  1.18  &  0.91\\
FU Ori$^{\dagger}$              &  0.59  &  0.88  & \, &  0.67  &  0.89\\
V1057 Cyg$^{\dagger}$           &  0.65  &  0.93  & \, &  0.89  &  0.92\\
\hline
\end{tabular}
\noindent
\flushleft
$^{\dagger}$Determined from Welty et al. (1992).  
$^{\ddagger}$Determined from Hamann \& Persson (1989).
\end{table}
In the four high resolution spectra described above, the
Ca\,{\sc ii}\,(2) triplet emission lines at 8498, 
8542 and 8662\,\AA\ are (apart from H$\alpha$) the strongest lines. 
Following Hamann \& Persson (1992a, b) we shall use our data on 
these lines as a diagnostic tool to obtain some information about the 
extended envelope around Z~CMa.
The profiles of this triplet in the spectra of Feb. 2000, Dec. 1991
and Dec. 1996 are given in Fig.~8.
It shows that they have P-Cygni shapes, as is characteristic for many
Herbig and T Tauri stars with outflow. As noted by Hamann \& Persson (1992a, b) 
other characteristics of this group are: (a) a tendency
for weaker emission lines to be narrower (in order H$\alpha$, Ca\,{\sc ii}\,(2), 
O\,{\sc i}, Fe\,{\sc ii}), (b) the Ca\,{\sc ii}\,(2) peak-flux  decreasing in order of increasing
wavelength and the EW of the 8542 Ca\,{\sc ii}  line being the largest of the
triplet EWs and (c) the presence of Fe\,{\sc i} emission lines.

In Table 5 we give the peak and EW ratios of the triplet lines for
the five spectra of Z~CMa and for those of FU Ori and V1057 Cyg. 
It is clear that the ratios of both FUors are different from those 
of Z~CMa and V645 Cyg. Our main argument that the observed Ca\,{\sc ii}\,(2) emission
is produced by the B star component of Z~CMa is that the EWs of the
Ca\,{\sc ii}\,(2) emission lines strongly increase with the increasing continuum
contribution by the B-star (10\%, 25\%, 59\%). However, in the
spectrum of 1996, where the contribution of the B0-star is only about
10\%, Fig. 8 shows a component in the red wing of the Ca\,{\sc ii} 
emission profiles, which could be due to the FUor component.
We also find a difference in the EWs of the 8540\,\AA\ and 8498\,\AA\  Ca\,{\sc ii}\,(2)
emission lines (both $-$3.08\,\AA) from the ESO CAT observations, 
obtained on Dec. 15, 1996, and those ($-$4.95 and $-$4.76\,\AA) observed with 
the WHT on Dec. 26, 1996. This is consistent with a smaller FWHM 
(93 km~s$^{-1}$ for both lines) on Dec. 15 than those from Dec. 26 
(97 and 101 km~s$^{-1}$). These differences could be caused by a variation 
in the disc of the B-star and/or in the FUor disc.

   In our spectra of 2000 and 1991 the (extinction corrected) ratio of
the total flux in the Ca\,{\sc ii} doublet
at 8912.7\,\AA\ and 8927.4\,\AA\ to the total flux of the Ca\,{\sc ii}\,(2) triplet is
0.036. Ferland \& Persson (1989) have estimated that in the temperature
range 5,000--10,000 K, (where Ca\,{\sc ii} is the dominating ionization stage)
it then follows that $n_e$ $\le$  10$^{11}$ cm$^{-3}$.  Because the EW ratio of
the triplet members is quite different from the ratio 0.11:1.0:0.56  
for the values of $\log (gf)$ (which should be the ratio if the lines
were optically thin) it has been recognised (Polidan \& Peters 1976)
that the Ca\,{\sc ii} triplet emission lines are optically thick and
saturated. Hamann \& Persson (1992b) showed that for $\tau$ = 1  a minimum
value for the Ca\,{\sc ii} column density can be given by 
$N$(Ca\,{\sc ii}) = 2.5 $\times$ 10$^{15}$ $\times$ ($v_{\rm turb}$/(50
km~s$^{-1}$)) cm$^{-2}$ provided that the line-width is determined by turbulence.
For Z~CMa this means that $N$(Ca\,{\sc ii}) $\ge$ 2.5 $\times$ 10$^{15}$ cm$^{-2}$ and therefore
$N$(H) $\ge$ 1 $\times$ 10$^{21}$ cm$^{-2}$. The condition for saturation is 
$\tau \times n_e$ $>$ 10$^{13}$ cm$^{-3}$, so that the
optical thickness of the line formation region should be $\tau$ $\ga$ 100 and
therefore $N$(Ca\,{\sc ii}) and $N$(H) much larger.

Hamann \& Persson (1992b) estimate the size of the Ca\,{\sc ii} emitting
region ($R_{\rm Ca \,{\sc ii}}/R_\star$) from the value of the surface flux of the 8542\,\AA\ 
Ca\,{\sc ii} line, defined by ($L_{\rm line}/L_\star) \sigma T_{\rm eff}^4$. If this 
line surface flux just equals
 $\pi \times \Delta v \times Bv(T_{\rm ex})$, where $\Delta v$ is the full width 
at half max. of the line and $T_{\rm ex}$ is the excitation temperature,
they assume that the line emitting region just covers the stellar
surface, while larger surface fluxes correspond with in proportion
larger surfaces. The authors have calculated the latter function for 
$T_{\rm ex}$ = 5000 K,
 $v$ = 25 and 250 km~s$^{-1}$ and  $T_{\rm ex}$ = 10,000 K and  
$v$ = 250 km~s$^{-1}$ and compared (their Fig. 9) the  line
surface fluxes, derived from the observations of the 8542\,\AA\ line for
various Herbig and T Tauri stars,
with the three calculated limits. For their data of Z~CMa (26--28 Dec.
1987) they assumed that it is a single
star with spectral  type F5. Then the derived 8542\,\AA\ line surface flux
is equal to the $T_{\rm ex}$ = 5000 K, $v$ = 250 km~s$^{-1}$ limit, 
so that the Ca\,{\sc ii} emitting region just covers a star with this $T_{\rm ex}$.

  From our new data we derived that the Ca\,{\sc ii} emission is formed in 
the neighbourhood of the B0 star component of Z~CMa. We can scale the line
surface flux from that of Hamann \& Persson (1992b) by noting that in Jan. 2000 
our estimate of the extinction corrected ($R_{\rm cs}$ = 4.2) continuum flux at
8542\,\AA\ for the B star component is 0.7 times the corresponding flux  of
the single F5 star in Dec. 1987.
On the other hand the EW of the 8542\,\AA\ line in Jan. 2000 is 3.4 times its
value, measured in Dec. 1987 and
the  $T_{\rm eff}$ of a B0III star ($\sim$ 30,000 K) is a factor 4.4 larger than
that of an F5I ($\sim$ 6900 K) star.  In this way 
we obtain for Jan. 2000 a line surface flux  of 1.71 $\times$ 10$^{10}$
erg cm$^{-2}$ s$^{-1}$.  This implies
a radius of $\sim$ 29.2 R$_\star$.  Fig. 10 of Hamann \& Persson (1992b) then shows
that this radius corresponds to
the Str\"omgren radius for a B0 star if $n_e$ $\sim$ 8 $\times$ 10$^{10}$ cm$^{-3}$, 
which is consistent with the value of $n_e$
estimated from the Ca\,{\sc ii} doublet and triplet observations above.
Ferland \& Rees (1988) and Ferland \& Persson (1989) 
have made calculations of the photoionization in the circumstellar
region and have shown that at high $n_e$
the Ca\,{\sc ii} emission can be produced just beyond the Str\"omgren radius.
Although in this region  the dominating ion is Ca\,{\sc iii}, the emissivity
of Ca\,{\sc ii} appears to be  high there because of the high temperatures.
 For early type B stars the emission of Ca\,{\sc ii} can arise from extended
high density regions from close to
the photosphere up to the Str\"omgren radius, e.g. in and around
circumstellar discs.

 We can obtain an estimate of the radius of the Ca\,{\sc ii} disc in Nov./Dec.
1991 by noting that for this epoch the EW of the 8542\,\AA\ line is 2.2 times 
smaller and the extinction corrected ($R_{\rm cs}$ = 4.2) continuum flux near 
the line is 3.4 times smaller than the corresponding values for Jan./Feb. 2000.  
This means that the line surface
flux in Nov./Dec. 1991 will be 0.13 times the line surface flux in
Jan./Feb. 2000, so that (if $T_{\rm ex}$  is the same)  it becomes 
2.3 $\times$ 10$^{10}$ erg cm$^{-2}$ s$^{-1}$, which implies an outer 
disc radius of $\sim$ 10.7 R$_\star$ in Nov./Dec. 1991.
In the same way we can use the data of Dec. 1996. For this epoch we
estimated that the observed continuum
flux of the B0 component is about 10\% of the total observed flux.
From this we derive that the extinction
corrected ($R_V$ = 4.2) flux near the 8542\,\AA\ line is 4.17 $\times$ 10$^{-12}$ 
erg cm$^{-2}$ s$^{-1}$, which is 2.3 times smaller than the
corresponding flux in Jan./Feb. 2000. The EW of the 8542\,\AA\ emission
in the Dec. 1996 spectrum is 3.7 times 
smaller than the corresponding EW in the Jan/Feb. spectra, so that
the line surface flux in Dec. 1996  $\sim$ 2.0 $\times$ 10$^{9}$ erg cm$^{-2}$ s$^{-1}$ 
and (if $T_{\rm ex}$ is the same) the outer disc radius will be $\sim$ 9.7 R$_\star$ in Dec. 1996.
We conclude that if  $T_{\rm ex}$ did not change, the outer radius of the
Ca\,{\sc ii}\,(2) emission region increased during the `outburst' of 
Jan 2000 with a factor of roughly 3. If  $T_{\rm ex}$
rises the increase in the outer radius will be smaller.

\subsection{The Fe\,{\sc ii} emission line spectra}
\subsubsection{Permitted Fe\,{\sc ii} lines}
\begin{figure}
\vspace*{0.15cm}
\centerline{\psfig{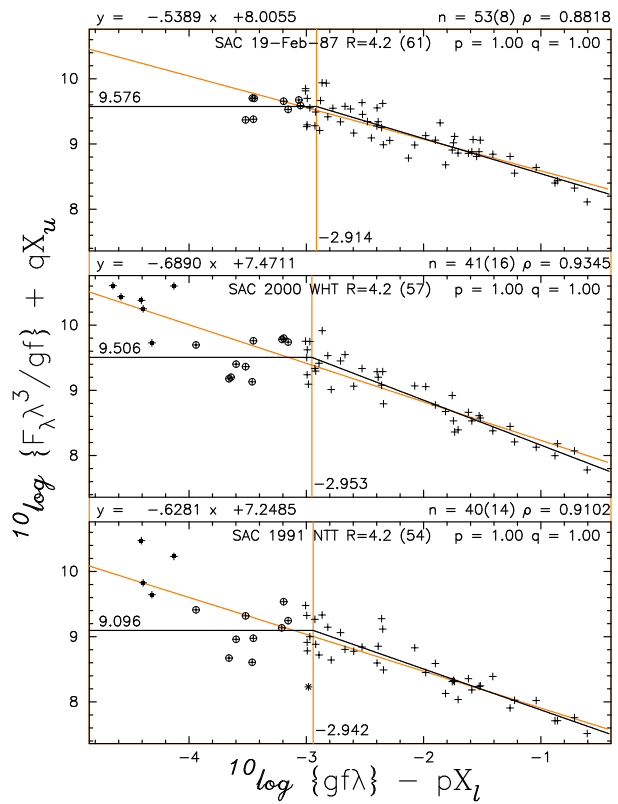}}
\caption[]{Self-absorption curves (SACs) for Fe\,{\sc ii} lines 
in the Z~CMa spectra from Feb. 1987 (top), Feb. 2000 (middle), 
and Dec. 1991 (bottom).  The + symbols indicate the optically 
thick Fe\,{\sc ii} lines, the $\oplus$ symbols indicate the group of 
optically thin Fe\,{\sc ii} lines, and the thick drawn lines 
are linear least square fits to each of these groups. The 
thin drawn line is a linear least square fit to both Fe\,{\sc ii} 
line groups.  Typical observational errors are comparable to 
the size of the plot symbols.  The thick crosses ({\bf +}) in 
the upper left corners of the middle (SAC 2000) and lower panel 
(SAC 1991) mark the positions of the forbidden [Fe\,{\sc ii}] 
lines.  They are not included in the least square fits.  The 
caption of each panel lists the number of optically thick lines 
($n$) included in the least square fits, and the correlation 
coefficient of these fits ($\rho$).}
\end{figure}
\begin{figure}
\vspace*{0.15cm}
\centerline{\psfig{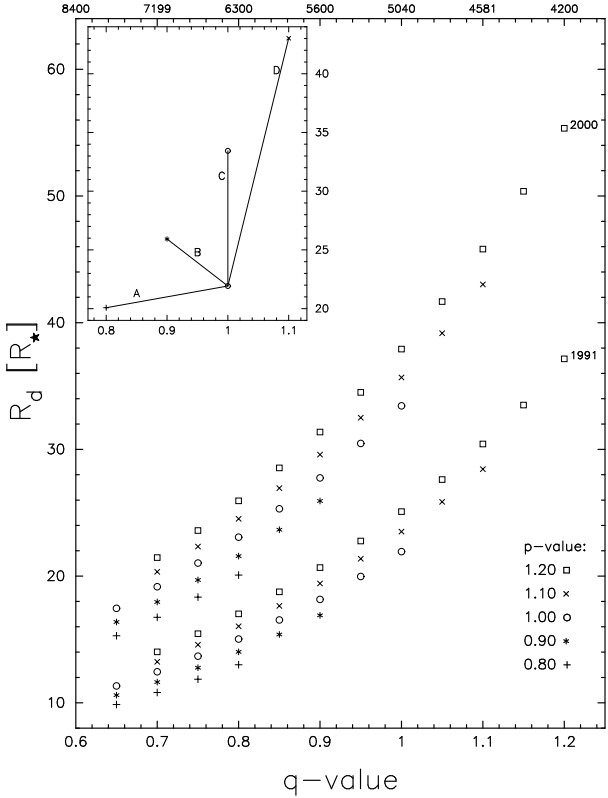}}
\caption[]{Z~CMa disc radii as a function of excitation 
parameter $q$ for the Feb. 2000 (top) and 
Dec. 1991 (bottom) Fe\,{\sc ii} spectra.  The insert shows the 
direction of varying excitation parameters, as discussed in 
the text.}
\end{figure}
The EWs of the Fe\,{\sc ii} emission lines in the spectra of Feb. 1987,
Nov/Dec. 1991 and Jan/Feb. 2000 
have been measured and are listed in Table 3 and 4.  We analysed
these line spectra in a statistical way:
the self-absorption curve (SAC) method (see Appendix), which is a
Curve of Growth version for emission lines. Strictly speaking we 
should correct the emission EWs for the line absorption, which
underlies each emission profile. This can be done for a number of
lines, which are observed in absorption in
spectra of low activity phases (Finkenzeller \& Jankovics 1984; Covino
et al. 1984; Welty et al. 1992).
However, for the majority of the lines these absorption components are
weak and difficult to estimate accurately. In addition part of the 
absorption components at low brightness phases may be contributed 
by the FUor component in Z~CMa.  Therefore we have neglected this 
correction for all lines.

Because the emission lines contribute very little (at most a few per
cent) to the continuum, we can use the
extinction-corrected continuum fluxes derived in the photometric
analysis in Sect. 3.2 to transform the EWs into absolute emission line
fluxes $F_\lambda$. By plotting for each line: $\log (F_\lambda \lambda^3/(gf)) + q \chi_u$
versus $\log (gf\lambda) - p \chi_l$, with $\chi_l$ and $\chi_u$ the excitation 
energies of the upper and lower terms of the transitions, 
we obtained a SAC curve for each of the three spectra. 
From the `bending point' we can obtain a  lower limit of the
Fe\,{\sc ii} column density and an upper limit of the radius of the Fe\,{\sc ii}
emission region in each case. In order to calculate $\log (N_0/g_0)$ we need
the value of $v_0$ in a thin disc, for which we take
the thermal velocity, which is about 2.0 $\times$ 10$^{5}$ cm s$^{-1}$.
In the Appendix it is explained how the values of $p$ and $q$ can be
obtained from a comparison between
the multiplets with common lower terms and of those with common upper
terms respectively.

In general the number of lines in a multiplet observed in our spectra
was about three, but for several multiplets such as 27, 28, 37, 38, 49,
73 and 74  we observed more lines, up to 8 or 9. The resulting values
of $p$ and $q$ show a certain spread related with the choice of the
multiplet pairs. This is partly caused by uncertainties in the EW 
measurements but perhaps also due to deviations from a thermal excitation 
mechanism.  For the 1987 spectrum we find $p$ = 0.9--1.13, 
$q$ = 0.92--1.05, for the 2000 spectrum $p$ = 0.8--1.10 and for
the spectrum of 1991 $p$ = 0.7--1.15, $q$ = 0.83--1.10. 
Since the uncertainties on these numbers are $\sim$20\%, 
it seems justified to assume that there is no significant difference
between the ($p$, $q$) values of the  three spectra and that the values of
$p$ and $q$ are close to 1. 
If the excitations of the upper and lower terms follow Boltzmann
distributions with (excitation) temperatures defined by 5040/$p$ for the
lower terms and 5040/$q$ for the upper terms of the transitions, we
expect $p$ to be somewhat larger than $q$: perhaps as large as 20\% , so
that for $p$ = 1 ($T_l$ =5040 K), $q$ could be 0.8 ($T_u$ = 6300 K).
With these choices for p and q we have calculated the SAC curves for
the three spectra (e.g. Fig.~9 for $p$ = $q$ = 1)
We note that the slope of the optically thick part of the curve (at the
right side of the bending point),
determined by least squares fitting, is around 0.6. Theoretical
predictions of the SAC curve for various models of line-formation
(Friedjung \& Muratorio 1987, Figs. 4c and 4f) indicate that such slope
values are consistent with line-formation in winds, confined to a
disc.

We derived the upper limits for  the disc radius and the lower limits
of the column densities of  Fe (or $N$(H)) for a range of values of 
$p$ and $q$, close to 1.0. Fig.~10 gives a survey of the resulting 
relative disc radii ($R_d/R_\star$) for the 1991 spectrum, and for the 
2000 spectrum for $R_{\rm cs}$ = 4.2 (for $R_{\rm cs}$
= 3.1  the values of $R_d/R_\star$ are smaller by a factor of 2). 
Note that, 
although the exact size of the Fe\,{\sc ii} emitting region in 1991 and 
2000 may be uncertain by a factor 2 or 3 (depending on the assumptions 
made about the excitation parameters), the ratio of the 
size of the Fe\,{\sc ii} emitting region in 1991 and 2000 has a much smaller 
uncertainty, perhaps $<$ 50\%. 
For ($p$, $q$) = (1, 1) when both excitation temperatures are 5040 K,
the upper limits of $R_d/R_\star$ for 1991, 2000 and 1987 are 22, 33 and 34.
This corresponds to the direction C in the insert of Fig.~10. 
For the 8542A Ca\,{\sc ii}\,(2) emission region we have found (previous section)
that  the Ca\,{\sc ii}\,(2) emission region seems to expand from 
$R_{\rm Ca\,{\sc ii}}/R_\star$ = 12 to $R_{\rm Ca\,{\sc ii}}/R_\star$ = 29 between 1991 and 2000. 
However, if we allow ($p$, $q$) to become lower between  1991 and 2000 the
radius will increase less because of the rising upper term excitation
temperature (directions B for (0.9, 0.9) and A for (0.8, 0.8)). It
is also possible that the excitation temperature decreases between
1991 and 2000 (direction D for (1.1 to 1.1)).
In case D the excitation temperature has decreased by 10\%,
but the radius of the Fe\,{\sc ii} disc increases by a factor of 2.0 
between 1991 and 2000.  For each value of ($p$, $q$) the SAC analysis 
also provides a lower limit to the column density of Fe\,{\sc ii} and 
(with the partition function and the $N$(Fe)/$N$(H) abundance ratio) 
a corresponding column density of Hydrogen.  We used the abundance 
ratio for the solar environment.  For (1, 1) we find 
$N$(H) $>$ 7 $\times$ 10$^{23}$ cm$^{-2}$ from the spectrum of 1991. 
In the points A, B, C and D of the spectrum of Jan./Feb. 2000 (Fig.~10) 
the ratios of $N$(H) with respect to the value of $N$(H) in (1, 1) 
from the 1991 spectrum are 0.28, 0.52, 1.03 and 2.06, respectively.  
The corresponding ratios of the disc radii upper limits are 0.92, 
1.18, 1.52 and 1.96.  In this example it seems that the column 
densities $N$(H) increase faster than the disc radii.  Since the 
bending point $X_0$ of the SAC is independent of continuum flux, 
the column density $N$(H) derived from it will not change for a 
different choice of spectral type.

\subsubsection{The forbidden Fe\,{\sc ii} lines}
In the spectra of Nov/Dec. 1991 and Jan./Feb. 2000 a small number of
[Fe\,{\sc ii}] emission lines
with $\log (gf)$ $>$ $-$7.9 have been found. The line list of Feb. 1987
(Hessman et al. 1991)  does not contain such lines, but in the original 
spectra used by Hessman et al. such lines can also be found.
The observed emission fluxes for these lines were treated in the same
way as the permitted Fe\,{\sc ii} lines and marked by filled plot symbols 
in Fig.~9.  The positions of these points are to the left and above the 
bending points of the curves for the permitted lines. Since these 
[Fe\,{\sc ii}] lines are optically thin they are not formed in
a dense region such as a disc, but probably in a extended region around the
disc. In order to make some estimate of the extension of this region we 
applied the modified form of the SAC ordinate:
$\log (V/d^2) =  Yf - \log(n_0/g_0) + 16.977$ in which $Yf$ is the normalized
emission flux of the line and
$n_0$ is the ground level volume density, which can be estimated from the
surface level density and
radius, determined from the permitted lines (Baratta et al. 1998).

The results are presented in Table 6, which shows that (for $p$ = 1.0,
$q$ = 1.0) the formation region of the [Fe\,{\sc ii}] lines is 
somewhat more extended than the disc, also in the plane of the disc. 
In the outburst state of Jan. 2000 it has expanded together with the 
disc. For the strongest [Fe\,{\sc ii}] lines the excitation region extends
out to nearly three times the upper limit of the disc radius. This result
appears to be valid for other values of $p$ and $q$ as well.
\begin{table}
\centering
\caption{Radii of [Fe\,{\sc ii}] emitting regions in the 1991 and 2000
spectra for ($p$,$q$)=(1.0,1.0).  $R_d$ = the disc radius.  Typical 
errors in these numbers are $\sim$ 20\%.}
\tabcolsep0.11cm
\small
\begin{tabular}{@{}llcccccc}
\hline
       &            & 4385\,\AA\ & 4815\,\AA\ & 5158\,\AA\ & 5261\,\AA\ & 5333\,\AA\ & 7155\,\AA\ \\
\hline 
 1991  &  $R/R_d$   &    --      &   2.9     &  2.7      &  1.8      &    --      &  1.5\\
 2000  &  $R/R_d$   &    2.3     &   2.0     &  2.3      &  1.8      &    2.0    &   1.2\\
\hline
\end{tabular}
%\noindent
%\flushleft
\end{table}

\subsection{The forbidden emission lines of O\,{\sc i}, Ca\,{\sc ii}, N\,{\sc i}, N\,{\sc ii} and S\,{\sc ii}}
There exist at least six recent profiles of the [O\,{\sc i}] 6300.3\,\AA\ emission 
in Z~CMa: the profile of March 31, 2002, a profile, obtained on Jan 23, 
1998 (Garcia et al. 1999), one profile from the WHT
spectrum of Dec. 1996, 4 profiles from Jan. 14 to 17, 1997 (Chochol et
al. 1998), two profiles from Dec. 20 and 21, 1991  and one profile from
Feb. 12, 1987 (Hessman et al. 1991). Fig.~11 shows the profiles for five 
dates.

Garcia et al. (1999) were able to separate the profiles from the 
primary (Herbig B0e) and secondary (FUor) star. The deconvolved image 
in their Fig. 3 indicates that the contribution of
the primary to the flux in the central peak of the [O\,{\sc i}] emission is
at least 2.5 times the contribution of the secondary (FUor) component of 
Z~CMa. This is confirmed by the fluxes of the recovered
spectra in their Fig. 2. In addition they showed that the shoulder at
the blue side of the line is  due to 
a small (mini-)jet, with a length of $\sim$1\arcsec\ in approximately the
direction of the large-scale outflow (Poetzel et al. 1989) and originating 
from the primary.  Garcia et al. suggest that this mini-jet was ejected 
at the time of the outburst of the primary in Feb. 1987.
\begin{figure}
\vspace*{0.15cm}
\centerline{\psfig{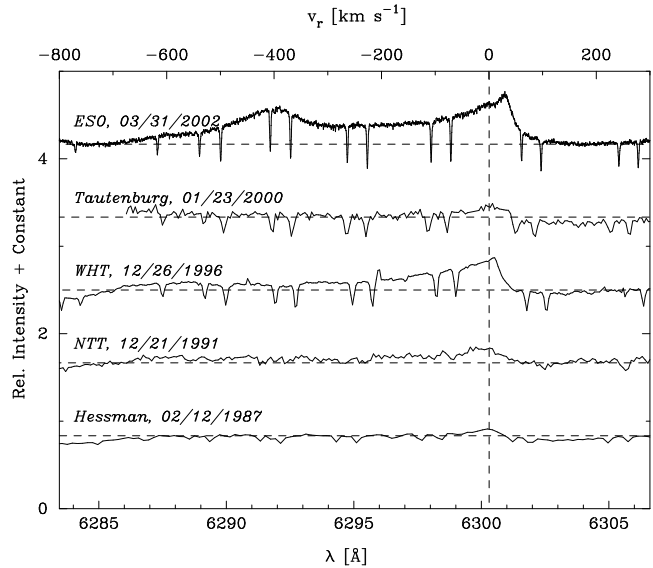}}
\caption[]{Variations in [O\,{\sc i}] 6300.2\AA\ line profile of 
Z~CMa between Dec. 1991 and Feb. 2002.  Dashed horizontal and vertical 
lines again indicate the continuum level and the rest-wavelength.  
The normalization of the profile of 2000 is somewhat uncertain 
because it is located near the edge of the echelle order.  
The narrow absorption lines seen in all spectra are due to telluric 
water vapour absorption, for which no correction was applied.}
\end{figure}

In our new [O\,{\sc i}] spectra, we note a high-velocity component, 
extending blueward of the main [O\,{\sc i}] emission peak.  In the 
spectrum of 1987 we only observe a single O\,{\sc i} peak without 
shoulder.   In all subsequent [O\,{\sc i}] 6300\,\AA\ profiles
a shoulder is present, starting at a distance of $\sim$ $-$600
km~s$^{-1}$ from the peak at 6300.3\,\AA.  

The most recent profile of the [O\,{\sc i}] line, taken on March 31, 2002, 
shows a remarkable increase in the flux of peak and shoulder. The appearance 
of these broad emission features suggests the sudden development of 
the high velocity outflow component.  
We also note the appearance of a second emission peak, centered on 
$\sim$ 6292~\AA, superimposed on the [O\,{\sc i}] 6300\,\AA\ line profile 
of Z~CMa.  This is similar to the second emission peak around 6294~\AA\ in 
the profile of the [O\,{\sc i}] 6300\,\AA\ line in the spectrum 
of PV Cep (Fig.~1 of Corcoran \& Ray 1997).  This second peak indicates 
a discrete high velocity component in the outflow, perhaps due to a change 
in collimation of a jet.  An alternative mechanism could be the emission 
of a discrete line or group of lines.  Plausible 
lines present in the 6290--6305\,\AA\ range are the lines of N\,{\sc ii}, 
with groupings of lines at 6286--6294~\AA\ and 
from 6298-6304~\AA. 

Corcoran \& Ray (1997) have noted that a high velocity component in
the [O\,{\sc i}] 6300\,\AA\ profile is a rare phenomenon for 
Herbig stars. It has only been observed for four stars:
V645 Cyg, Z~CMa, PV Cep
and LkH$\alpha$ 233.  These stars are young, embedded stars which are
associated with stellar jets and molecular outflows and 
probably accreting rapidly. The jets form close to the stellar
surface and are ejected along the rotation axis of the star. The
collimation of such jets within
150--200 AU has been successfully modeled with a stellar or 
disc-generated wind in a rapidly
rotating magnetosphere around such stars. The fact that such high
velocity jets are not observed for the majority of the Herbig stars 
may be caused by differences in the evolution of the collimating magnetic field.

We have also inspected the [Ca\,{\sc ii}] 7291.46\,\AA\ and 7329\,\AA\ profiles 
in the Tautenburg spectrum and in the spectra of Nov./Dec. 1991 and Dec. 1996.
It is remarkable that they are similar to the central peak of the [O\,{\sc i}]
6300\,\AA\ line, but they clearly lack the extended blue shoulder. The same
seems to hold for the [O\,{\sc i}] 6363\,\AA, the [S\,{\sc ii}] 6731\,\AA, 
the [N\,{\sc ii}] 6583\,\AA\ and the [O\,{\sc i}] 5577\,\AA\  profiles, but 
this is more difficult to verify since the flux in the peak of these lines 
is three times lower than that of the [O\,{\sc i}] 6300\,\AA\  line.
It is not completely clear whether the central peaks of the [N\,{\sc ii}] and
[S\,{\sc ii}] lines are mainly due to the primary B0e star. However, it 
seems convincing that the total flux is dominated by the primary since 
the [N\,{\sc ii}] and [S\,{\sc ii}] fluxes seem confined to the jet direction, 
which runs through the primary (Poetzel et al. 1989; Garcia et al. 1999).
The weakness of the [N\,{\sc ii}] and [S\,{\sc ii}] lines does not allow 
us to obtain reliable profiles.
If most of the forbidden lines originate in the neighbourhood of
the primary, we can try to use the flux ratios to obtain a rough 
estimate of the conditions in the far outer atmosphere of the B0e star.
Hamann (1994) calculated ratios of various forbidden line fluxes with
respect to the line
flux of [O\,{\sc i}] 6300\,\AA\  under the assumptions of coronal ionization
equilibrium and solar abundances.
This allows us to estimate $T_e$ and $n_e$ from the flux ratios of 
[N\,{\sc ii}]\,(6583\,\AA), [S\,{\sc ii}]\,(6731\,\AA),
[Ca\,{\sc ii}]\,(7291\,\AA) and [O\,{\sc i}]\,(5577\,\AA) to [O\,{\sc i}]\,(6300\,\AA).
We have used the observed EWs and intrinsic continuum fluxes after
circumstellar extinction corrections with $R_{\rm cs}$ = 4.2.
From the diagrams of Hamann we then obtain for $T_e$ = 13,000 K the
values of $n_e$ in the formation region of the various lines (see Table 7).
\begin{table}
\centering
\caption{$n_e$ (cm$^{-3}$) for  $T_e$ =13,000 K. Typical errors in these
numbers are $\sim$ 30\%. }
\tabcolsep0.11cm
\small
\begin{tabular}{@{}lcccc}
\hline
                 &         1996          &         1991         &       2000            &      1987\\
                 &     ($V$ = 10.2)      &      ($V$ = 9.6)     &    ($V$ = 9.2)        &   ($V$ = 8.7)\\
\hline
$[$S\,{\sc ii}] 6732\,\AA  &  8 $\times$ 10$^{4}$  &  5 $\times$ 10$^{4}$ & 2.0 $\times$ 10$^{4}$ & 1.5 $\times$ 10$^{4}$\\
$[$Ca\,{\sc ii}] 7291\,\AA &      --               &      --              & 2.0 $\times$ 10$^{6}$ &   --\\
$[$O\,{\sc i}]  5577\,\AA  &  3 $\times$ 10$^{6}$  &       --             & 2.5 $\times$ 10$^{6}$ &   -- \\
$[$N\,{\sc ii}] 6583\,\AA  &    --                 &  3 $\times$ 10$^{6}$ & 3.0 $\times$ 10$^{6}$ & 2.2 $\times$ 10$^{6}$\\
\hline
\end{tabular}
%\noindent
%\flushleft
\end{table}

For $T_e$ = 15,000~K we find the same trends: (a)  for a fixed $T_e$ the
[S\,{\sc ii}] line comes from the
lowest density and therefore at the largest distance from the stellar
envelope.
(b) the densities decrease when the Be star becomes brighter. This suggests 
an expansion of the outer region where the forbidden lines are formed.

\subsection{The IRTF spectrum}
\begin{figure}
\vspace*{0.15cm}
\centerline{\psfig{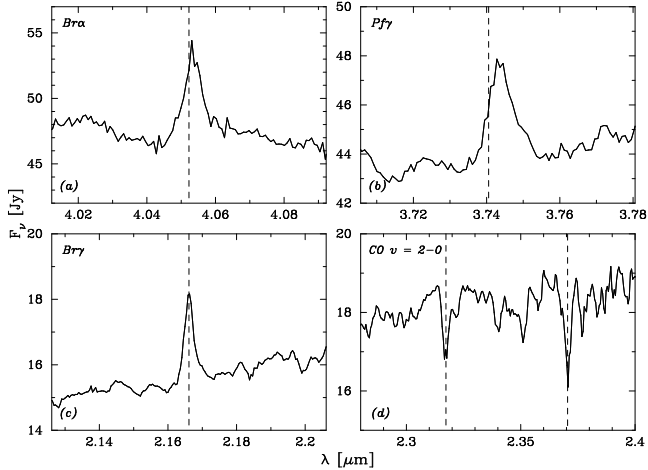}}
\caption[]{Fragments of the Z~CMa IRTF spectrum of Dec. 2001, 
showing the lines of Br$\alpha$ (4.05~$\mu$m), Pf$\gamma$ (3.74~$\mu$m), 
Br$\gamma$ (2.67~$\mu$m) and the CO band-head around 2.3--2.4 $\mu$m.}
\end{figure}
\begin{table}
\centering
\caption{Lines in the IRTF spectrum}
\tabcolsep0.13cm
\small
\begin{tabular}{@{}clccc}
\hline
$\lambda$ [$\mu$m] & \multicolumn{1}{c}{Ident.} & EW [\AA] & FWHM [km~s$^{-1}$] & $v_{\rm out}$ [km~s$^{-1}$]\\
\hline
4.05   & Br$\alpha$  & $-$7.75  & 370  & $-$1600\\
3.74   & Pf$\gamma$  & $-$6.37  & 545  & $-$1670\\
2.37   & CO (2--0)   & ~~~4.28  & 328  & --\\
2.32   & CO (2--0)   & ~~~2.31  & 328  & --\\
2.17   & Br$\gamma$  & $-$5.00  & 375  & $-$1500\\ 
\hline
\end{tabular}
%\noindent
%\flushleft
\end{table}
In the IRTF spectrum of Dec. 2001 we observe the emission lines 
Br$\alpha$ (4.05 $\mu$m), Pf$\gamma$ (3.74 $\mu$m), Br$\gamma$ (2.16 $\mu$m), 
He\,{\sc i} (2.06 $\mu$m), Br$\delta$ (1.94 $\mu$m) and absorption 
in the CO $v$ = 2--0 overtone bands around 2.3~$\mu$m.  
Br$\beta$ is in the observed wavelength-range, but it is obscured by terrestrial 
atmospheric bands. Fig. 12  shows the profiles of Br$\alpha$, Pf$\gamma$, 
Br$\gamma$ and the CO band.  The EWs are given in
Table 8. The CO absorption most likely arises in the FUor component 
of Z~CMa (Hartmann et al. 1989).  The infrared H\,{\sc i} emission 
lines appear to have similar outflow velocities as their optical 
counterparts (Fig.~1); they may be related to the inner accretion disk 
around the Herbig B0 primary.  
Muzerolle et al. (1998b) have derived empirical relationships
between the luminosity of Br$\gamma$ and the accretion rate of 
T Tauri stars. Our continuum flux at 2.2 $\mu$m corresponds to 
$K$ = 4.03\mag. We therefore have a Br$\gamma$ emission line flux of
6.0 $\times$ 10$^{-12}$ erg~cm$^{-2}$~s$^{-1}$ or $\log F$(Br$\gamma$) = $-$11.2. 
From Fig. 3 in  Muzerolle's paper we then find by extrapolation that the 
accretion rate of Z~CMa in Dec. 2001 should be 
$\sim$2 $\times$ 10$^{-5}$ M$_\odot$~yr$^{-1}$.

\subsection{Summary}
In the preceding sections we have compared the spectra of the
`outburst' phases in Feb. 1987 and Jan./Feb. 2000 with those of 
the `quiescent' phases Nov./Dec. 1991 and Dec. 1996.
The most remarkable distinction between these phases is the appearance
of deep absorption lines of He\,{\sc i} at 7065, 6678 and 5876\,\AA\ 
and of O\,{\sc i} at 7773\,\AA\ and 6156\,\AA\ in the outburst spectra.  
We interpret this appearance as a clear indication that the `outburst' 
occurs in the primary Be component of Z~CMa. 

The strong similarity between the spectra of Z~CMa with those of the
class of Herbig Be stars suggests
a similar interpretation: the emission lines are formed in a
circumstellar disc. Because of the direction of the huge bi-polar jet
(large radial velocities in the knots),  the association of this
outflow with the primary star (Garcia et al. 1999) and the observed
polarization of the lines (Whitney et al. 1993), we assume that we are
seeing the disc under an angle of at most 40 degrees.

Following Hamann \& Persson (1992b) we have used the fluxes of the strong
Ca\,{\sc ii}\,(2) triplet to estimate the extent of the corresponding
excitation region. We found that the outer radius of this region expands 
from 10 R$_\star$ in 1991 and 1996 to $\sim$ 29 R$_\star$ during the outburst in
Jan./Feb. 2000. The line profiles of Ca\,{\sc ii}\,(2) may indicate that during
the faintest phase (Dec. 1996) a small contribution of Ca\,{\sc ii} emission
from the FUor component may be present. The analysis of the Fe\,{\sc ii}
emission multiplets shows a similar expansion of 
the Fe\,{\sc ii} region. This expansion is probably
somewhat smaller than that of Ca\,{\sc ii}: 23 R$_\star$ in 1991 to 32 R$_\star$ in 2000,
but 37 R$_\star$ in Feb. 1987. [Fe\,{\sc ii}] emission extends up to twice this
distance. Other forbidden lines in the spectra such as [S\,{\sc ii}],
[N\,{\sc ii}] and [Ca\,{\sc ii}] correspond to very low values of $n_e$ 
and are formed much further away, probably in the bipolar outflow 
(Poetzel et al. 1989).

The similarity of the Balmer and Paschen outflow velocity pattern
with those of the Ca\,{\sc ii}\,(2) and Fe\,{\sc ii}\,(42) lines (both groups have several
outflow components with velocities up to 450--600 km~s$^{-1}$) suggests that
these lines are formed in the same region. The emission components of
Ca\,{\sc ii} (8542\,\AA), P11 (8863\,\AA) and
O\,{\sc i} (8446\,\AA) are very similar in profile. Especially in the 1991 spectrum, 
where the blue absorption wings are less conspicuous, the profiles are very similar 
to those of the T Tauri stars BP Tau and UY Aur.
Muzerolle et al. (1998) have modeled these BP Tau
profiles with a nonrotating,
axisymmetric dipole field for magnetic accretion and gave rough
relationships between the extinction-free
line fluxes of the Ca\,{\sc ii} (8542\,\AA), O\,{\sc i} (8446\,\AA) and 
P11 on one hand and the accretion rate derived from
the UV/blue excess continuum (Gullbring et al. 1998) on the other. We
have used these relationships
to derive for Z~CMa accretion rates of $\log \dot{M}$ (M$_\odot$~yr$^{-1}$) = $-$4.4 for
the O\,{\sc i} and P11 and $-$5.3 for the Ca\,{\sc ii}
emission lines by extrapolation of Fig. 9 of Muzerolle et al. (1998).
From the flux of Br$\gamma$ and Fig. 3 in the paper of Muzerolle et
al. (1998b) we obtain by extrapolation that $\log \dot{M}$ (M$_\odot$~yr$^{-1}$) = $-$4.7.
It therefore seems possible that in Z~CMa these lines show inflow
(magnetospheric accretion) as well as outflow.
 
Because of the changes in line emission flux and outflow velocities 
we expect that both flows are different in the various phases of Z~CMa.
The He\,{\sc i} absorption lines have high excitation energies and must
therefore be formed close to the star while
the O\,{\sc i} absorption lines probably have their origin in an 
extended envelope above and below the disc plane.
Most of the emission lines of  O\,{\sc i}, Ca\,{\sc ii}, Fe\,{\sc ii}, Fe\,{\sc i}, 
Cr\,{\sc ii}, Mg\,{\sc ii} and the Paschen lines form in the disc.
Na\,{\sc i}\,D, K\,{\sc i} and Ca\,{\sc ii}\,K are formed further 
out in the envelope. The Ca\,{\sc ii}\,K absorption pattern shows similarities 
with those of the Na\,{\sc i}\,D lines.
These lines  may receive contributions from the envelope of the FUor
component as well. However, it seems highly probable that the
variations in these profiles also reflect changes in the outer
envelope of the primary Be star during the outburst phases.

Two new phenomena have been observed in connection with the outburst
of Jan. 2000:
the appearance of a strong Fe\,{\sc ii} 9997\,\AA\ emission line and of 
an extended blue shoulder in the [O\,{\sc i}]
6300\,\AA\ emission profile in 2002. The origin of the first emission is
not yet clear. It is possible that Ly$\alpha$
fluorescence is the mechanism for its production. The emission
profiles of the [O\,{\sc i}] 6300\,\AA\ and the [Ca\,{\sc ii}]
7291, 7323\,\AA\ doublet profiles show a gradual growth in EW since
Dec. 1987. The blue shoulder in the [O\,{\sc i}] 6300\,\AA\ profile, 
detected two years after the outburst of  Jan. 2000 is possibly caused 
by a sudden change in the collimation process of the mini-jet, which 
could have been ejected in Feb. 1987.

\section{Discussion}
The spectrum of Z~CMa in its normal state has been
satisfactorily explained by Hartmann et al. (1989),
and Welty et al. (1992) in terms of that of a rotating, optically thick, 
accretion disc. From a total luminosity of 
3000 L$_\odot$ Hartmann et al. (1989) derived for 
$v \sin i$ $\sim$ 100 km~s$^{-1}$ an inner disc radius of 16 R$_\odot$ 
and an accretion rate of $\sim$ 10$^{-3}$ M$_\odot$~yr$^{-1}$ to a central 
star of 1--3 M$_\odot$ and radius $\sim$ 9--16 R$_\odot$. However, during 
the outburst of Feb. 1987 (Hessman et al. 1991) the spectrum was completely
different and the observers of this high
state could not explain it in terms of the same or an adapted model.
It were the spectropolarimetric observations of Whitney et al. (1993)
that showed that the usually small contribution of the infrared primary
star to the visual continuum could rise to $\sim$66\% of the total 
continuum flux during the outburst of 1987.
Whitney et al. (1993) have noted that in this high state the emission
spectrum of Z~CMa is very similar
to that of the Herbig Be star MWC 1080. The agreement is not limited
to the emission lines but also to the profile of the Na\,{\sc i}\,D 
line (Finkenzeller \& Mundt 1984). Similar to Z~CMa, the
He\,{\sc i} lines at 6687 and 5876\,\AA\ of MWC 1080 are in absorption 
(Yoshida et al. 1992).

  We also note a strong spectral similarity  with the luminous young
object V645 Cyg (Hamann \& Persson 1989), which has been classified as
O8.  Especially  the broad Fe\,{\sc ii} 9997\,\AA\ emission line and 
the broad O\,{\sc i} (7774\,\AA) absorption line together with a 
relatively  weak  O\,{\sc i} (8446\,\AA) emission is very similar to 
the high state of Z~CMa.  This indicates an important role of Ly$\alpha$ 
and Ly$\beta$ emission in both stars.
A difference is that the He\,{\sc i} (6678\,\AA) P-Cygni profile of 
V645 Cyg has a more prominent emission component
than its counterpart in Z~CMa. This may point to a somewhat hotter
inner disc. Another feature which
both objects have in common is the high velocity component of the 
[O\,{\sc i}] 6300\,\AA\ emission profile (see below).

The spectrum of Z~CMa has also been compared with the rich emission
line spectrum of V380 Ori.
However,  the spectrum of V380 Ori (Rossi et al. 1999) shows some
marked differences with those
of Z~CMa and MWC 1080. Its He\,{\sc i} lines at 6678 and 5876\,\AA\ 
are generally in emission, but occasionally
in absorption. It has been suggested by Rossi et al. and others that
this may indicate the
presence of a chromosphere or a hot accretion spot close to the
stellar surface. Its O\,{\sc i} (7773\,\AA)
line is also in emission, which is not unusual for various early A-type
Herbig stars, such as AB Aur (A0e), HD 163296 (A2e) and 
BD +61\degr154 (B8e) (Felenbok et al. 1988). An independent argument for
a classification of V380 Ori as a B9--A0e star is the cut-off of the
ultraviolet continuum near 1300\,\AA\ 
in the low resolution IUE spectra of V380 Ori (Rossi et al. 1999).

    The spectral type of the Z~CMa primary component was found to be
close to B0e (Sect. 4.2).
From $V_0$ = 6.05\mag\ and a distance of 1050 pc we find an absolute
magnitude $M_V$ = $-$4.1\mag, which
may indicate a classification B1--2 III. In Sect. 3.2 we have derived the
luminosities of the primary
component for Nov. 21, 1991 (the date of the polarization spectra of
Whitney et al. 1993) and
Feb. 20, 1987 (the `high state' of Z~CMa). The results for three
assumptions for $R_{\rm cs}$ (the ratio of total to specific absorption 
of the circumstellar envelope): 3.1, 4.2 and 6.0 are given in Table 2.

     Is may be interesting to compare the temperatures and the de-reddened 
luminosities of this table with recent calculations of the 
pre-main sequence evolution of young massive stars by Behrend \& Maeder (2001).
The comparison shows that in the high state for $R_{\rm cs}$ = 3.1 and 
$R_{\rm cs}$ = 4.2 the emission line star can be
identified as a pre-main sequence star, since then it can be modelled
as a 16 M$_\odot$ B0 III star ($T_{\rm eff}$  
$\sim$ 31,600 K) on the birthline with an age of 3 $\times$ 10$^{5}$ yr 
and a luminosity of  47 L$_\odot$ and radius 7.6 R$_\odot$.
For $R_{\rm cs}$ = 6.0 the luminosity is too high and the position of the
star is above the birthline and, in order to be observable, it should be 
a post main sequence star. From the post main sequence
evolutionary calculations of Schaller et al. (1992) we find that it
could be 20 M$_\odot$ B0 III star
with an age of  $\sim$ 6.6 Myr. This is close to the estimated ages of the
nearby B0e stars HD 53367 and HD 53755 (paper II).  On the other hand 
the parameters of the inner star of the FUor component,
estimated by Hartmann et al. (1989), correspond to a 
3 M$_\odot$ star with a radius of 7 R$_\odot$,
an effective temperature of 5000 K, a luminosity of 28 L$_\odot$ and 
an age of 3 $\times$ 10$^{5}$ yr. Its position in
the HR diagram is just below the birthline. In this case both
components of Z~CMa have the same age and could have been born
together.

    Recently Yorke \& Sonnhalter (2002) have published evolutionary
models for star formation in which radiative acceleration of dust
grains has been taken into account in the radiative transfer. The
results of the evolution depend on the opacity law of the dust. For a
frequency-dependent opacity law due to a mixture
of two dust components with specified size distributions the evolution
leads to the formation of massive stars via disc accretion with a
massive bipolar jet. For a `gray' opacity law, however, the evolution
proceeds without bipolar jet and results in a thin, disc-like object.
Since in these models the birthline is raised by a factor ten (for B0
stars) with respect to the previous models, 
they can easily interpret the Z~CMa primary luminosity of Feb. 1987
with $R_{\rm cs}$ up to 6 as the luminosity of
a pre-main sequence star. The paper shows 2-dimensional  models of the
immediate neighbourhood of the stars for a wide grid of masses, ages 
and luminosities. 

For our case of 3.1 $\times$ 10$^{5}$ L$_\odot$ (Table 2, 
$R_{\rm cs}$ = 6) we are not far from a model of a 38 M$_\odot$ star, 
close to the birthline with an age of 4 $\times$ 10$^{4}$ Myr (model
F120, 11c). For this age the model predicts a mass accretion rate smaller 
than 10$^{-4}$ M$_\odot$~yr$^{-1}$, which is consistent with our estimate 
from the 
Br$\gamma$ flux in Sect. 4.8.  Similar to Z~CMa, the 2-dimensional model 
shows an indication of an accretion disc  and of a strong bi-polar outflow. 
The dynamical time scale of 2--3 $\times$ 10$^{4}$ years for the `slow' knots 
(Poetzel et al. 1989) in the bipolar jet fits within the model age of this 
B0e star.  

For ages comparable with those of the B0 star models the FUor
component could correspond to a `gray' opacity model such as G60 of
Yorke \& Sonnhalter (2002) but for a lower luminosity and mass than
calculated in their Table 4. Such a model does not develop a bipolar
outflow, but evolves to a disc-like structure, perhaps
a FUor. In this case we may have a simultaneous birth of the Z~CMa
components.  This is in agreement with the polarization results of
Fischer et al. (1998), who came to the conclusion that both 
components must have had a common formation history. However, 
the models by Yorke \& Sonnhalter imply a difference in grain properties 
for the regions in which the B0 and the FUor component are formed. 
Because the present rotation velocity of the B0 component and its
previous evolution are not known we cannot be certain of the
evolutionary  age, but so far it seems that the age estimates of Z~CMa
with both models are much lower than those of the other massive stars
in the CMa~R1 complex, which are closer to the main sequence (paper
I, II). How can the low age of Z~CMa be fitted in the evolutionary
history of CMa~R1?   The formation must be related to fragmentation in
the high density in the dark cloud S296 where Z~CMa is situated, but so
far  it remains unclear which process initiated the contraction in the
cloud which started the recent star formation.

Apart from the long-term evolution of Z~CMa there remain a number of
questions concerning the recent behaviour of the two components: 

(1) The cause of the photometric variability:
At least two types of variabilities have been observed: 
(a) irregular low amplitude variations, which
seem on the average to follow density variations in the circumstellar
dust. V.S. Shevchenko made an extensive search of periodicities in the 
$UBVR$ photometry, collected by the ROTOR program during the
years 1981--1998. A period of 393 days was found with an amplitude of
0.3\mag\ in $U-B$ and 0.08\mag\ 
in $B-V$. The underlying cause of such a periodicity is not yet known
but pulsational instability may be a possibility.
(b) Strong and fast rises in brightness (outbursts) around 1968 ($\Delta V$ 
$\sim$ 0.7\mag), 1987 ($\Delta V$ $\sim$ 0.6\mag) and 2000 
($\Delta V$ $\sim$ 1.2\mag). We have now some indications that the second 
outburst has caused the ejection of the 
mini-jet emitting [O\,{\sc i}] 6300\,\AA\ (Garcia et al. 1999), 
similar to those observed in other young emission-line objects such as 
PV Cep, V645 Cyg and LkH$\alpha$ 233 (Corcoran \& Ray 1997).
The third outburst (in Jan. 2000) probably caused a change in the
[O\,{\sc i}] 6300\,\AA\ profile (see Fig.~11) which could be due to 
a change in the collimated outflow. The
outbursts therefore could be related to changes in the upper
layers of the stellar interior, which
entail changes in the magnetic field configuration. Such changes could
have an internal origin or be induced
by an origin from outside the star, e.g. by tidal interaction during
approaching component stars.  The latter possibility (Bonnell \& Bastien 1992) 
can be ruled out by the following argument: The 
projected distance between the components of Z~CMa is $\sim$ 0.1\arcsec\, 
which at a distance of 1050 pc corresponds to $\sim$105 AU. This is seen 
in projection and therefore the minimum distance of the components. 
A minimum for the period can then be estimated from Kepler's law. For 
$M_1$ = 38 M$_\odot$ and $M_2$= 3 M$_\odot$, we find a minimum period 
of 168 yrs.  This is too large for an observed
recurrence of outbursts at a timescale of once per 10--12 yrs. 
The remaining possibilities are that the primary star has a, as of yet 
undetected, close binary component or that the outbursts are induced 
by variations in the star itself.

(2)  According to Garcia et al. (1999) the direction of the mini-jet 
coincides with that of the large bi-polar outflow of Z~CMa and the 
outflow is driven by the massive B0 component. Also the evolutionary models
of Yorke \& Sonnhalter (2002) predict large bi-polar outflows during
the formation of massive stars.
However, Vel\'azquez \& Rodr\'{\i}guez (2001) claim that the large bipolar
outflow coincides with the radio-jet through the FUor component. If this 
is true, the question remains how the cavity in the cocoon is formed, through
which we observe the optical spectrum of the B-star. One possibility is
that it has been formed in the past when the infrared source approached the FUor. 
Somehow the direction of the bipolar outflow is then conserved in the 
infrared source and transferred to the mini-jet.

\section*{Acknowledgments}
The authors are indebted to Drs. H. van Winckel and G. Meeus for 
obtaining the Dec. 1996 spectrum of Z~CMa at the WHT, and to 
Dr. Lee Hartmann for sharing the original spectra of Z~CMa during 
the 1987 outburst. 
We also would like to thank the service-observers at the ING group 
(especially Dr. I. Skillen and Dr. B. Garcia) for their excellent 
work in obtaining the Feb. 2000 WHT service-mode observations of Z~CMa, 
and the observers of the ROTOR photometric program at Mt. Maidanak 
for their quick notification of the outburst of Dec. 1999. 
We also would like to thank Dr. L.N. Berndnikov for obtaining 
$UBVRI$ photometry of Z~CMa during his stay at the South African 
Astronomical Observatory in March 2002.  
One of us (H.T.) thanks in particular Dr. R. Viotti and Dr. M. Friedjung 
for helpful discussions concerning the application of the SAC method. 
We also thank the referee, Dr. B. Reipurth, for his constructive 
remarks which improved content and presentation of the manuscript.  
This research has made use of the Simbad data base, operated at CDS, 
Strasbourg, France.  Atomic data used in this study were extracted 
from the Atomic Line List maintained at 
http://www.pa.uky.edu/$\sim$peter/atomic/.

\appendix
\section{The self-absorption curve (SAC) method}
The observation of many multiplets of emission lines of Fe\,{\sc ii}, 
Ti\,{\sc ii} and Cr\,{\sc ii} (Table 4) allows us
to derive some information concerning the formation regions of these
lines with the help of an
emission curve of growth method. The method is based on the analysis of
the self-absorption curve (SAC) which was developed by 
Friedjung \& Muratorio (1987) and has been successfully
applied to various Fe\,{\sc ii} emission line spectra (e.g. 
Muratorio et al. (1992) for the case of KQ Pup)\footnote{An 
independent derivation of the SAC method has been given by 
Kastner (1999)}. A manual for
the use of the method has been written by Baratta et al. (1998).

The SAC curve describes the relation between the line strength and
optical depth of emission lines
in a multiplet. This relation can be empirically determined by plotting
$\log (F_\lambda \lambda^{3}/gf)$ (where $F_\lambda$ is the
normalised absolute line flux and $gf$ is the oscillator strength of the
line) versus $\log (gf)$ (which is the
optical depth at line-center up to an additional constant) for the
various lines of each multiplet.
In these empirical plots the lines through the points of each multiplet
should be parallel to each other
and to the still unknown theoretical SAC function $Q(\tau)$, which should be
zero for $\log \tau$ $<$ 0  and should
should decrease approximately linearly for $\log \tau$ $>$ 1 . The emission
lines in the horizontal part of the
SAC (with $\log \tau$ $<$ 1) are optically thin, whereas those in the 
decreasing part are increasingly optically
thick.  For the various Fe\,{\sc ii} emission lines the oscillator strength and
excitation energies have been taken from the tables of Kurucz (1981).

The various curves for multiplets with common upper terms\footnote{In the 
visual part of the spectrum Fe\,{\sc ii} multiplets with 
common upper terms are: 27, 38, 48 and 73 ($^4P(^5D)^4D$),  28, 32, 37, 49 and 55 
($^4P(^5D)^4F$),  26,42 and 36 ($^4P(^5D)^6P$), 29 and 74 ($^4P(^5D)^4P$), 
and 41 and 46 ($^4P(^5D)^6F$).} will be parallel and shifted by $p \chi_l$ in
horizontal direction with respect to each other, whereas those with
common lower terms\footnote{In the visual part of the spectrum Fe\,{\sc ii} 
multiplets with common lower terms are: 27, 28 and 29 ($D^6S(^3P)^4P$),  
36, 37 and 38 ($D^6S(^3F)^4F$),  40, 41, 42 and 43 ($D^5S^2(^7S)^6S$), 
46, 48 and 49 ($D^6S(^3G)^4G$) and 72, 73 and 74 ($D^6S(^3D)^4D$).} will be shifted 
by $q \chi_u$ in vertical direction with respect to each other. Here $\chi_l$
and $\chi_u$ are the excitation potentials of the lower and upper  terms in
the transitions and $p$ and $q$ are the values of $5040/T_l$ and
$5040/T_u$, where it is assumed that the levels have Boltzmann-type
populations with excitation
temperatures $T_l$ and $T_u$ for the lower and upper terms respectively.
Since we know the excitation
potentials, we can determine the values of $p$ and $q$ from the horizontal
and vertical shifts and therefore the Boltzmann temperatures.

By plotting $Y = \log (F_\lambda \lambda^{3}/gf) + q\chi_u$ versus 
$X = \log (gf\lambda) - p \chi_l$ for the lines of each multiplet, we can
obtain one common empirical SAC curve for the Fe\,{\sc ii} lines. 
The value $X_0$ on the X-axis, at which the
empirical SAC curve starts to decrease for increasing $X$, gives us
$\log (N_0/g_0)$ up to constant $\log v_o + 1.576$
(in cgs units), where $v_o$ is the line broadening velocity and $N_0$ is
the ground state column density of Fe\,{\sc ii} from which the total 
Fe and H column densities can be estimated. From the normalized emission 
line flux
$Y_0$ in this point $X_0$, where the empirical SAC curve starts to
decrease, it is then possible to estimate the
extension of the surface area $S$ of the emission region (perpendicular
to the line of sight) using 
$\log (S/d^{2})$ = $Y_0 - \log (N_0/g_0) - w + 16.977$), where $d$ is 
the distance to the source (here 1050 pc) and
$w = - (p-q) \times (\chi_l + \chi_u)/2$. An example of the empirical 
SAC curves of Fe\,{\sc ii} emission lines of three spectra
of Z~CMa (observed in Feb. 1987, Jan. 2000 and Nov. 1991) is given in
Fig.~9.

For spectra with only optically thick lines, one can only obtain the
decreasing part of the SAC curve, so that 
the values of $X_0$ and $Y_0$ have to be estimated. The method then only
gives a lower limit to the column density and an upper limit to the extension 
of the emission line region. On the other hand the method can be
applied to optical thin forbidden lines, to derive a characteristic
volume (or radius) of their emission region.
Finally the slope of the decreasing part of the theoretical SAC curve
has been shown to depend on the dynamical conditions in the emission 
region (Friedjung \& Muratorio 1987). In our case of Z~CMa the
slope  is close to $-0.6$, which may indicate that the Fe\,{\sc ii} 
lines are formed in a disc wind (Fig. 4f in Friedjung \& Muratorio 1987).

\bsp
\label{lastpage}
\end{document}